\documentclass[preprint]{aastex}

\received{}
\accepted{}
\journalid{}{}
\articleid{}{}
\slugcomment{To appear in {\it the Astrophysical Journal}}

\shortauthors{KOBAYASHI \& NAKASATO}
\shorttitle{Chemodynamical Simulations of The Milky Way Galaxy}

\def\gtsim {>\kern-1.2em\lower1.1ex\hbox{$\sim$}~}   
\def\ltsim {<\kern-1.2em\lower1.1ex\hbox{$\sim$}~}   

\begin{document}

\title{Chemodynamical Simulations of the Milky Way Galaxy}
\author{Chiaki KOBAYASHI$^1$ and Naohito NAKASATO$^{2,3,4}$}
\affil{$^1$ Research School of Astronomy \& Astrophysics, The Australian National University, Cotter Rd., Weston ACT 2611, Australia; chiaki@mso.anu.edu.au}
\affil{$^2$ School of Computer Science and Engineering, The University of Aizu, Aizu-Wakamatsu City, Fukushima, 965-8580, Japan}
\affil{$^3$ Center for Computational Sciences, University of Tsukuba, Tsukuba, Ibaraki 305-8577, Japan}
\affil{$^4$ Computational Astrophysics Laboratory, RIKEN, Wako, Saitama 351-0198, Japan}
\begin{abstract}
We present chemodynamical simulations of a Milky Way-type galaxy using a self-consistent hydrodynamical code that includes supernova feedback and chemical enrichment, and predict the spatial distribution of elements from Oxygen to Zinc.
In the simulated galaxy, the kinematical and chemical properties of the bulge, disk, and halo are consistent with the observations.
The bulge formed from the assembly of subgalaxies at $z\gtsim3$, and has higher [$\alpha$/Fe] ratios because of the small contribution from Type Ia Supernovae.
The disk formed with a constant star formation over $13$ Gyr, and shows a decreasing trend of [$\alpha$/Fe] and increasing trends of [(Na,Al,Cu,Mn)/Fe] against [Fe/H].
However, the thick disk stars tend to have higher [$\alpha$/Fe] and lower [Mn/Fe] than thin disk stars.
We also predict the frequency distribution of elemental abundance ratios as functions of time and location, which can be directly compared with galactic archeology projects such as HERMES.
\end{abstract}


\section{Introduction}

While the evolution of the dark matter
is reasonably well understood, the evolution of the baryonic component is much
less certain because of the complexity of the relevant physical processes, such
as star formation and feedback.
With the commonly employed,
schematic star formation criteria alone, the predicted star formation rates 
(SFRs) are higher than what is compatible with the observed luminosity
density. Thus feedback mechanisms are in general invoked to reheat gas and
suppress star formation.
We include the feedback from stellar winds, core-collapse supernovae (normal Type II Supernovae (SNe II) and hypernovae (HNe)), and Type Ia Supernovae (SNe Ia) in our hydrodynamical simulations.
Supernovae inject not only thermal energy
but also heavy elements into the interstellar medium (ISM),
which can enhance star formation.
Chemical enrichment must be solved as well as energy feedback.
Supernova feedback is also important for solving the angular momentum problem \citep{ste99,som03} and the missing satellite problem \citep{moo99}, and for explaining the existence of heavy elements in intracluster medium \citep{ren93} and intergalactic medium, and possibly the mass-metallicity relation of galaxies \citep[hereafter K07]{kob07}.

Since different heavy elements are produced from different supernovae with different timescales,
elemental abundance ratios can provide independent information on ``age'' \citep[e.g.,][hereafter KN09]{kob09}.
Therefore, stars in a galaxy can be used as fossils where the chemical enrichment history of the galaxy is imprinted, and this approach is called as galactic archeology.
In the Milky Way Galaxy, the detailed structures of kinematics and metallicities have been obtained with SDSS (the Sloan Digital Sky Survey) and RAVE (the Radial Velocity Experiment), and will be extended with Pan-STARRS and SkyMapper.
In the next decade, high-resolution multi-object spectroscopy (e.g., APOGEE with SDSS and HERMES on the AAT) and space astrometry mission (e.g., GAIA) will provide 6D-kinematics and elemental abundances for a million stars in the Local Group.
In order to untangle the formation and evolution history of the galaxy from observational data, a ``realistic'' model that includes star formation and chemical enrichment is required.

Although there are several simulations of the Milky Way-type galaxy with very high resolutions \citep[e.g.,][]{die07}, there are few studies with hydrodynamics and chemical enrichment \citep[e.g.,][]{sca09}.
In most of the previous work, however, the adopted assumptions are not appropriate for comparison with observations; the effect of mass-loss from stars is ignored, the timescales of SNe Ia is given by only one parameter, the nucleosynthesis yields of supernovae are outdated, and so on.
Recently, several new findings have been obtained from the collaboration between observations, stellar astrophysics, and galactic chemical evolution modeling.
From the observed light curves and spectra of nearby core-collapse supernovae, it has been shown that there are energetic supernovae, HNe, ejecting more than ten times larger explosion energy ($E_{51}\gtsim10$) and a certain amount of iron \citep[e.g.,][]{nom06} as well as $\alpha$ elements (O, Mg, Si, S, Ca, and Ti). 
Such HNe play a major role in reproducing the observed [Zn/Fe] trend in the solar neighborhood, where [Zn/Fe] is $\sim0$ over $-2\ltsim$ [Fe/H] $\ltsim0$ and possibly increases toward lower [Fe/H] \citep[hereafter K06]{kob06}.
From the observations of supernova rates in various types of galaxies \citep{sul06,man06}, it has been shown that there is a young population of SNe Ia in addition to the old population that is usually found in present-day elliptical galaxies.
The metallicity effect on the occurrence of SNe Ia, which has been proposed by \citet[hereafter K98]{kob98}, plays a major role in reproducing the observed [($\alpha$,Mn)/Fe]-[Fe/H] relations (KN09).

Including the up-to-date knowledge of chemical evolution of galaxies, we provide chemodynamical simulations of a Milky Way-type galaxy from CDM initial conditions.
In \S 2, we summarize our simulation code and chemical enrichment sources.
In \S 3, we show the differences between the chemical properties of the disk and bulge, namely the metallicity distribution function, the age-metallicity relation, and the [$\alpha$/Fe]-[Fe/H] relation, and predict the [X/Fe]-[Fe/H] relation.
We also propose a new diagram of [$\alpha$/Fe]-[Mn/Fe] as a sequence of the SN Ia contribution.
\S 4 gives our conclusions.

\section{Model}

\subsection{Hydrodynamics}
\label{sec:model}

The details of our GRAPE-SPH code are described in \citet{kob04} and can be summarized as follows.

\begin{figure}
\center 
\includegraphics[width=10cm,angle=-90]{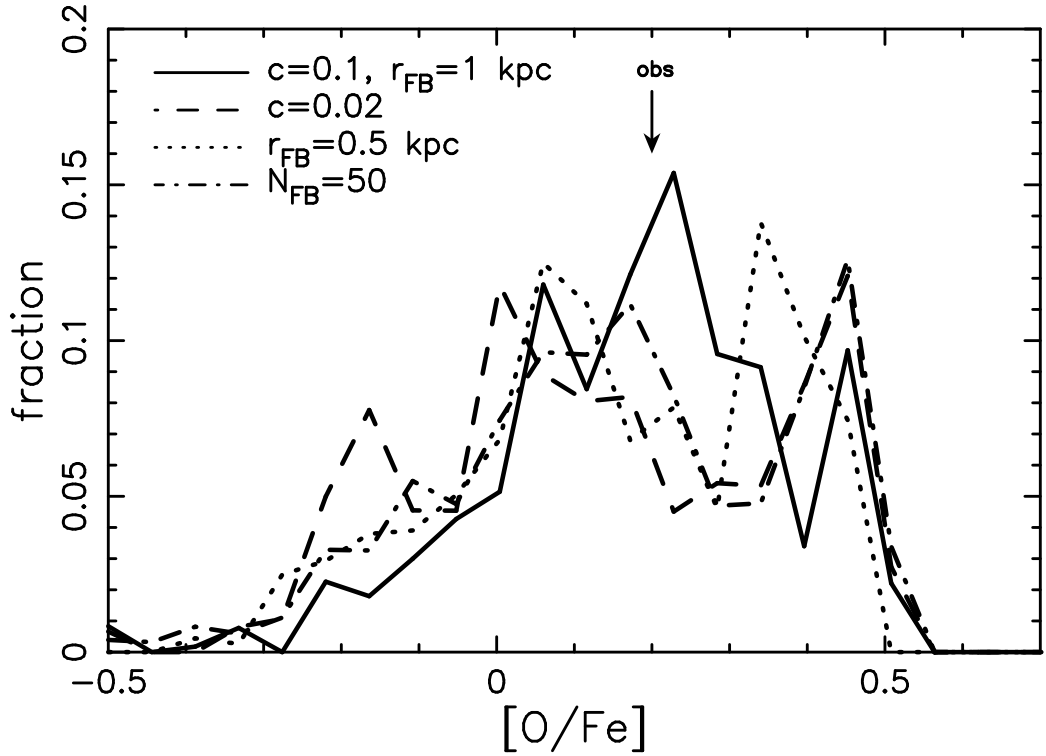}
\caption{\label{fig:afenum05}
Distribution functions of [O/Fe] for stars in the solar neighborhood ($z=0$) at $-0.5 \le$ [Fe/H] $\le -0.3$ for the standard model (solid line) and the models with the star formation timescale $c=0.02$ (dashed line), the smaller feedback radius $r_{\rm FB}=0.5$ kpc (dotted line), and the fixed feedback number $N_{\rm FB}=50$ (dot-dashed line).
The arrow shows the peak value of observations.
}
\end{figure}

\begin{enumerate}
\renewcommand{\labelenumi}{\roman{enumi})}

\item The gravity is computed with the tree method using the special purpose computer GRAPE (GRAvity PipE) system at the National Astronomical Observatory of Japan.
For hydrodynamics, the Smoothed Particle Hydrodynamical (SPH) method is adopted, and the SPH formulation is the almost same as \citet{nav93}.
The GRAPE-SPH code is highly adaptive in space and time by means of individual smoothing lengths and individual timesteps, and it has very high performance (a week for one simulation with $N\sim 50,000$).

\item {\bf Radiative cooling} is computed with the metal-dependent cooling functions, which are generated with the MAPPINGS III software \citep{sut93} as a function of [Fe/H].
[O/Fe] is fixed with the observed [O/Fe]-[Fe/H] relation in the solar neighborhood.

\item Our {\bf star formation} criteria are the same as \citet{kat92};
1) convergent, 2) cooling, and 3) Jeans unstable.
The SFR is determined from the Schmidt law;
the star formation timescale is proportional to the dynamical timescale ($t_{\rm sf} \equiv \frac{1}{c}t_{\rm dyn}$), where $c=0.1$ is chosen from the size-luminosity relation of elliptical galaxies \citep{kob05}.
In addition, we find that $c=0.1$ gives a better fit to the elemental abundance ratios in the Milky Way Galaxy than $c=0.02$ (see Fig. \ref{fig:afenum05}).
We also adopt the probability criterion \citep{kat92}.
In our simulations, the SFR depends both on the local density and metallicity, which is different from other simplified models such as one-zone models.

\item If a gas particle satisfies the star formation criteria,
a part of the mass of the gas particle turns into a star particle.
Since one star particle has the mass of $10^{5-6} M_\odot$,
the star particle is not a single star but an association of many stars.
The masses of individual stars span according to the {\bf initial mass function} (IMF). 
We adopt the Salpeter IMF that is invariant to time and metallicity with a slope $x=1.35$ for $0.07-120M_\odot$.

\item For the {\bf feedback} of energy and heavy elements, we do not adopt the instantaneous recycling approximation.
Via stellar winds, SNe II/HNe, and SNe Ia, thermal energy and heavy elements are ejected from an evolved star particle as a function of time, and distributed to the surrounding gas particles within a sphere of feedback radius $r_{\rm FB}$.
The mass transfer is weighted by the SPH kernel from the gas particle to the dying star particle.
This feedback scheme is better than that implementing a fixed number of feedback neighbor particles ($N_{\rm FB}$) in galaxy-scale simulations; the metallicity of gas particles does not depend very much on the local density if $N_{\rm FB}$ is fixed.
We find that $r_{\rm FB} = 1$ kpc gives a better fit to the elemental abundance ratios than $r_{\rm FB} = 0.5$ and $3$ kpc, or $N_{\rm FB} = 50$ and $100$ (see Fig. \ref{fig:afenum05}).
Note that the diffusion between gas particles is supposed to enhance the mixing of heavy elements but is not included in our model.

\item The {\bf photometric evolution} of one star particle is identical to the evolution of a simple stellar population (SSP).
SSP spectra are taken from \citet{kod97} as a function of age and metallicity.
The photometric evolution of the galaxy is calculated from the summation of the SSP spectra of star particles with various ages and metallicities.

\end{enumerate}

\subsection{Chemical enrichment}
\label{sec:chem}

The scheme to include the detailed chemical enrichment into SPH simulations has been proposed by \citet{kob04} and updated as follows.

{\bf Hypernovae (HNe)} ---
The explosion mechanism of core-collapse supernovae is still uncertain, although a few groups have presented feasible calculations in exploding core-collapse supernovae \citep{kit06,mar09,bru09}.
However, the ejected explosion energy and $^{56}$Ni mass (which decays to $^{56}$Fe) can be directly estimated from the observations, i.e., the light curve and spectra fitting of individual supernova.
As a result, it is found that many core-collapse supernovae ($M \ge 20M_\odot$) have more than ten times larger explosion energy ($E_{51}\gtsim10$) and produce a significant amount of iron \citep{nom06}.

We adopt our nucleosynthesis yields of core-collapse supernovae from K06,
depending on the progenitor mass, metallicity ($Z=0-0.05$), and the explosion energy (normal SNe II and HNe).
The progenitors mass range is $8-50M_\odot$ but the metal ejection is only from $10-50M_\odot$.
Assuming that a large fraction of supernovae with $M \ge 20M_\odot$ are HNe,
the evolution of the elemental abundance ratios from oxygen to zinc is in excellent agreement with observations in the solar neighborhood, bulge, halo, and thick disk.
In particular, to solve the [Zn/Fe] problem, the HN fraction $\epsilon_{\rm HN}$ has to be as large as $\sim 50\%$ (K06).
Note that Ti is underabundant at $-3\ltsim$[Fe/H]$\ltsim0$, which will be solved with the 2D calculation of nucleosynthesis \citep{mae03}.
The observed increasing trend toward the lower metallicity may suggest that higher energy under the assumption of inhomogeneous enrichment.

At high metallicity, since neutron-rich isotopes $^{66-70}$Zn are produced, $\epsilon_{\rm HN}$ can be as small as $1\%$.
In this work, we assume the metal-dependent efficiency of HNe ($M \ge 20M_\odot$), $\epsilon_{\rm HN}=0.5, 0.5, 0.4, 0.01$, and $0.01$ for $Z=0. 0.001, 0.004, 0.02$, and $0.05$\footnote{$\epsilon_{\rm HN}=0.5$ is adopted independent of metallicity in K06, K07, and KN09.}, which gives better agreement with the observed present HN rate \citep{pod04}.
Pair-instability supernovae, which produce much more Fe, more [S/Fe], and less [Zn/Fe], should not contribute in the galactic chemical evolution \citep{kob10b}.

{\bf Type Ia Supernovae (SNe Ia)} ---
The progenitors of the majority of SNe Ia are most likely 
the Chandrasekhar (Ch) mass white dwarfs (WDs).
For the evolution of accreting C+O WDs toward the Ch mass,
two scenarios have been proposed;
one is the double-degenerate scenario, i.e., the merging of double C+O WDs 
with a combined mass surpassing the Ch mass limit.
However, it has been theoretically suggested that it leads to accretion-induced collapse rather than SNe Ia \citep{nom91}, and the lifetimes are too short to reproduce the chemical evolution in the solar neighborhood (KN09).
The other is
the single-degenerate (SD) scenario,
i.e., the WD mass grows by accretion of hydrogen-rich matter via mass transfer from
a binary companion.
The mass accretion rate is very limited to trigger carbon deflagration \citep{nom82}, but the allowed parameter space of binary systems can be significantly increased by the WD wind effect if the metallicity is higher than [Fe/H] $\sim -1$ (K98).
Two progenitor systems are found for red-giant (RG) companions with initial masses of $\sim 1 M_\odot$ 
and near main-sequence (MS) companions with initial masses of $\sim 3 M_\odot$ \citep{hac99u}.

In our simulations, based on the SD scenario, the lifetime distribution function of SNe Ia is calculated with Eq.[2] in KN09, taking into account the metallicity dependence of the WD winds (K98) and the mass-stripping effect on the binary companion stars (KN09).
The lifetimes of SNe Ia are determined from the lifetimes of companion stars, and thus the lifetime distribution function shows double peaks;
one is for the MS+WD systems 
with timescales of $\sim 0.1-1$ Gyr, which are dominant in star-forming galaxies.
The other is for the RG+WD systems with $\sim 1-20$ Gyr timescales, which are dominant in early-type galaxies.
Although the metallicity effect has not been confirmed from the supernova survey yet, it is more strongly required in the presence of the young population of SNe Ia, to be consistent with the chemical evolution of the Milky Way Galaxy.
The absolute rates of each component depend on two binary parameters, which are determined by the requirement to meet the observed metallicity distribution function and [$\alpha$/Fe]-[Fe/H] relations in the solar neighborhood, and the observed supernova rates in various types of galaxies.
This SN Ia model gives much better agreement with the observed elemental abundance ratios (from O to Zn) than models with the double-degenerate scenario or metal-independent exponential lifetime functions (KN09).

For SNe Ia, we take the nucleosynthesis yields from \citet{nom97}, where the metallicity dependence is not included.
Ni is overproduced at [Fe/H] $\gtsim -1$, which will be solved by tuning the propagation speed of the burning front and the central density of the white dwarf \citep{iwa99}.

{\bf Stellar winds} --- 
Needless to say, the envelope mass and pre-existing heavy elements are returned by stellar winds from all dying stars.
In addition, Asymptotic Giant Branch (AGB) stars with initial masses between about $0.8-8 M_\odot$ (depending on metallicity) produce light elements such as C and N, while the contribution of heavier elements are negligible in the galactic chemical evolution \citep{kob09b,kob10a}.
Rotating massive stars ($\gtsim 40 M_\odot$) could also produce C and N \citep[e.g.,][]{mey02}, but are not included.

\subsection{Initial Condition}

\begin{deluxetable}{lcccc|ccc}
\tablenum{1}
\tablecaption{\label{tab:ic}
Initial and Final Quantities. (1) comoving radius, (2) initial number of all particles, (3) total mass, (4) mass of an SPH particle, (5) final number of all particles, (6) final baryon fraction within 100 kpc, and (7) final stellar fraction within 100 kpc.}
\footnotesize
\tablewidth{0pt}
\tablehead{
 & (1) & (2) & (3) & (4) & (5) & (6) & (7)\\
Model & $r_{\rm init}$ & $N_{\rm tot,init}$ & $M_{\rm tot}$ & $M_{\rm sph}$ & $N_{\rm tot,final}$ & $b$ & $f_*$
}
\startdata
Standard & $1.5$ Mpc & 55,534 & $0.8 \times 10^{12} M_\odot$ & $3.8 \times 10^6 M_\odot$ & 180,293 & 0.14 & 0.23\\
Wider region & $2$ Mpc & 117,525 & $1.7 \times 10^{12} M_\odot$ & $3.8 \times 10^6 M_\odot$ & 316,991 & 0.15 & 0.25\\
Higher resolution & $1.5$ Mpc & 209,133 & $0.8 \times 10^{12} M_\odot$ & $1.0 \times 10^6 M_\odot$ & 445,259 & 0.12 & 0.29\\
\enddata
\end{deluxetable}

The initial condition is generated in the same way as in \citet{nak03} and \citet[2005]{kob04}.
The CDM initial fluctuation is generated by the COSMICS package \citep{ber95}.
The cosmological parameters are set to be
$H_0=70$ km s$^{-1}$ Mpc$^{-1}$, $\Omega_m=0.3$, $\Omega_\Lambda=0.7$, $\Omega_{\rm b}=0.04$, $n=1$, and $\sigma_8=0.9$.
We generate a periodic boundary condition with the lattice size $8$ Mpc having a top-hat perturbation of amplitude $3\sigma$ in radius $1.4$ Mpc 
at the starting redshift of $z=24$.
Then, particles in a spherical region with a comoving radius that contains $\sim 10^{12} M_\odot$ are picked up.
Finally, because the simulated field is not large enough to generate tidal torque, the initial angular momentum is given to the system in rigid rotation with a constant spin parameter as large as $\lambda \equiv J|E|^{1/2}G^{-1}M^{-5/2} = 0.1$ \citep{nak03,kob03}.
Such a large value is required to reproduce disk galaxies, although rare in cosmological simulations \citep[e.g., ][]{war92}.

\begin{figure}
\center 
\includegraphics[width=12.5cm]{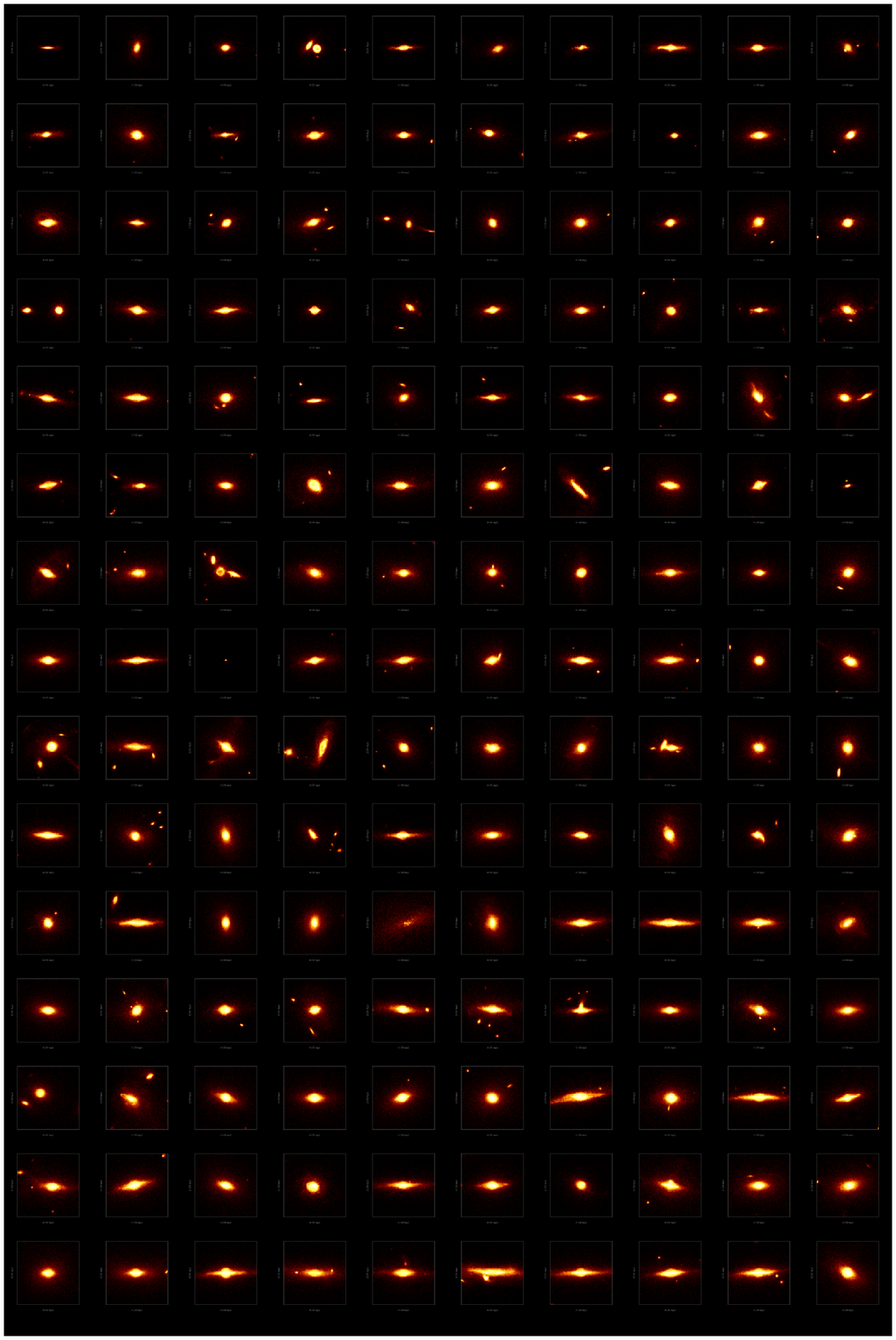}
\caption{\label{fig:ic}
Projected stellar density maps for the edge-on views at $z=1.28$ for 150 galaxies from various initial conditions.
}
\end{figure}

It seems not easy to form the Galactic disk in the CDM Universe
as suggested by previous works (e.g., \citealt{tot92,kau93}, but see \citealt{kod09,hop09}).
To form a disk galaxy, it is necessary that the system does not undergo the major mergers with mass ratios larger than 0.2 and at higher redshifts than $z=2$, which destroy the disk structure and the radial metallicity gradient \citep{kob04}.
Even in some cases without major mergers, disk structures cannot form \citep[e.g.,][]{sca09}, although this argument depends on the modeling of feedback and numerical resolution.
We have run simulations with 150 initial conditions until $z=1.28$ including gas dynamics and basic chemical enrichment (Fig. \ref{fig:ic}).
The disk structure with a scale length of $3-4$ kpc is seen in 48 runs, and Milky Way type galaxies form only in 5 runs, from which we select one realization (143rd in Fig. \ref{fig:ic}) for detailed simulations.

\begin{figure}
\center 
\includegraphics[width=9.5cm,angle=-90]{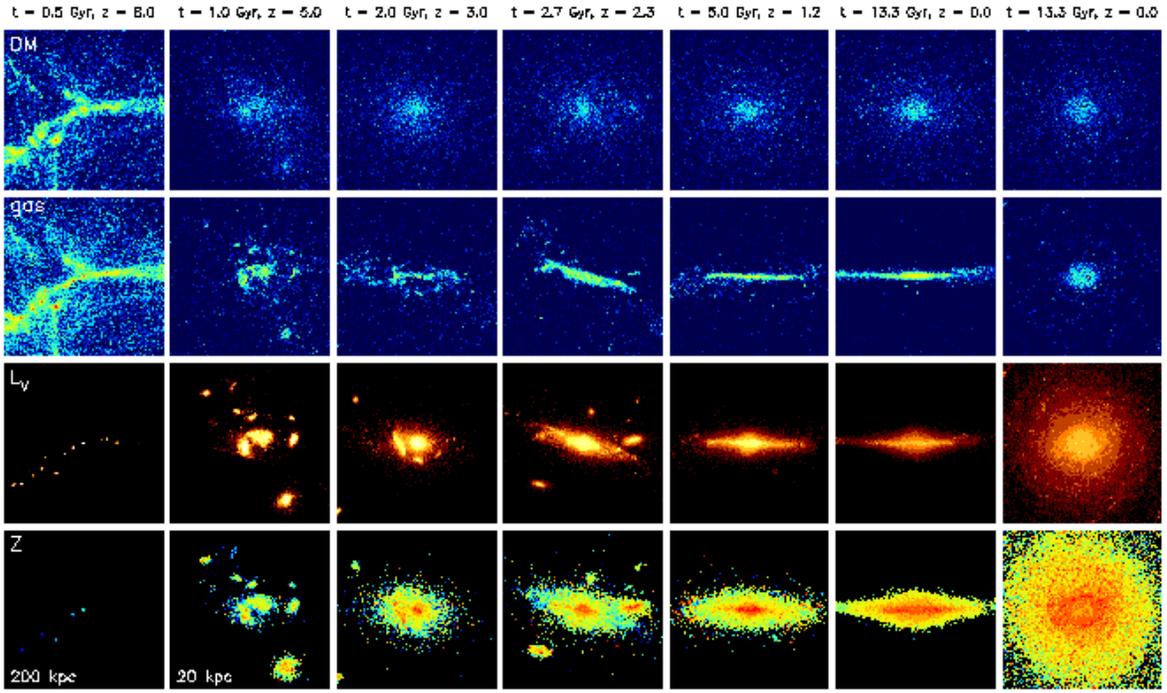}
\caption{\label{fig:map}
Time evolution of the projected density of dark matter (first row), gas (second row), V-band luminosity (third row), and the luminosity-weighted stellar metallicity (forth row) in the range of [M/H]$=-1$ to $0.1$ at $z=8, 5, 3, 2.3, 1.2$, and $0$.
The leftmost panels are 200 kpc and the other panels are 20 kpc on a side.
The rightmost panels are for the face-on views at $z=0$.
}
\end{figure}

\begin{figure}
\center 
\includegraphics[width=12cm,angle=-90]{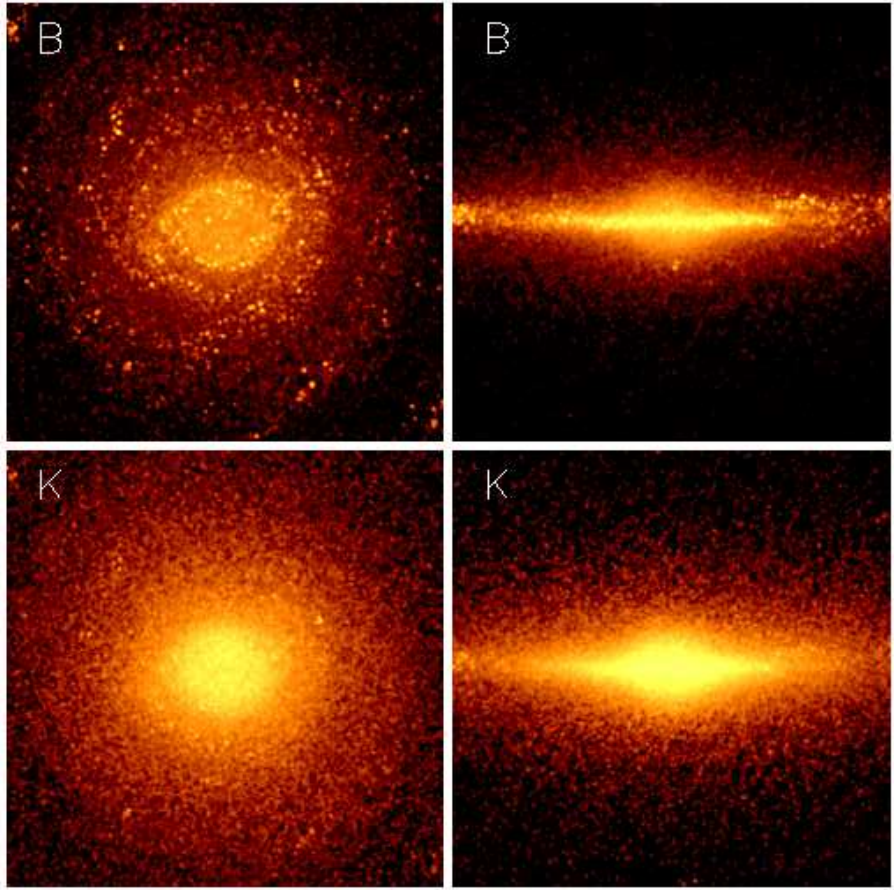}
\caption{\label{fig:map2}
B-band (upper panels) and K-band (lower panels) luminosities for the edge-on (left panels) and face-on (right panels) views at $z=0$ in $20$ kpc on a side.
}
\end{figure}

\begin{figure}
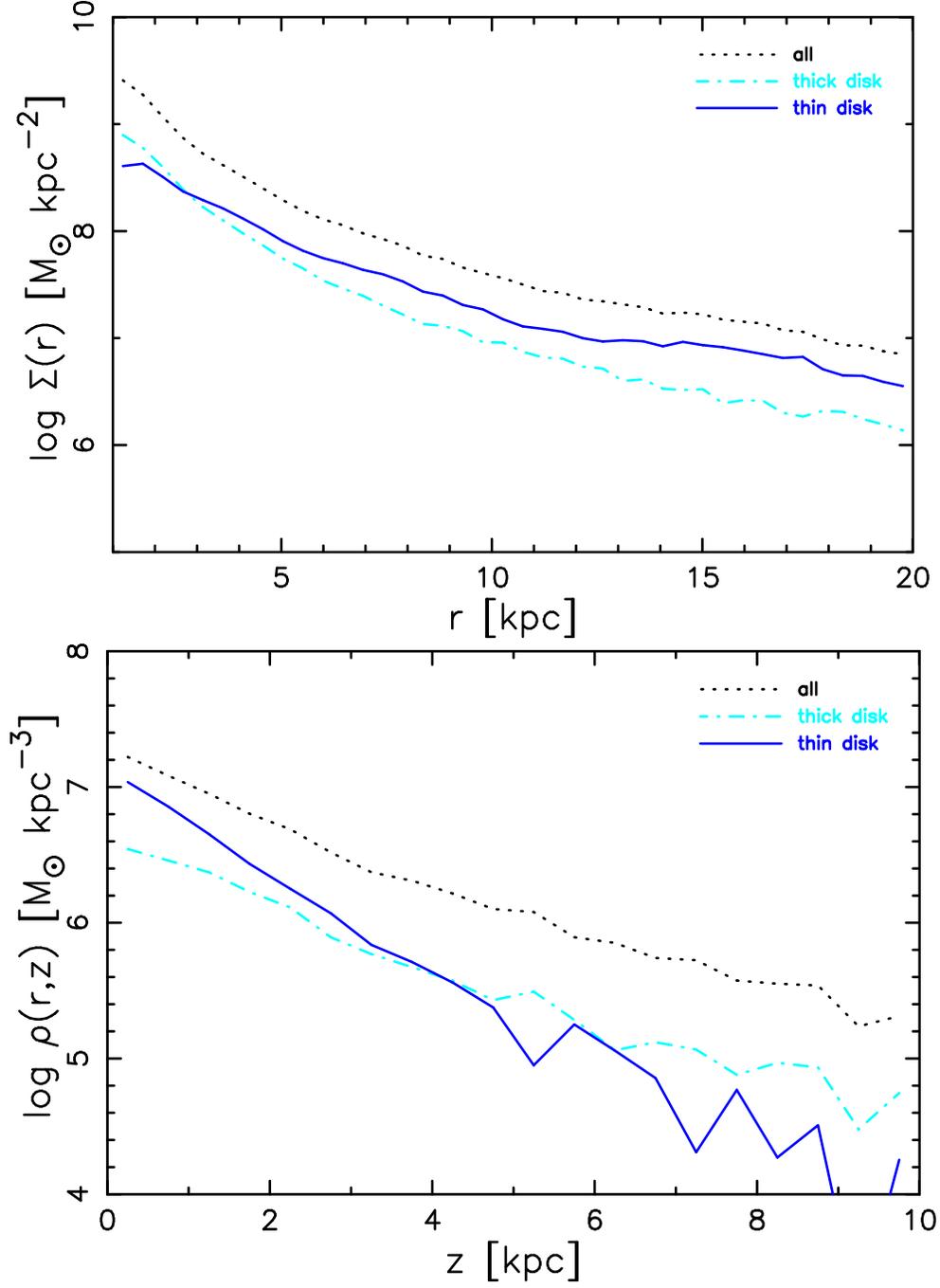

\center 
\includegraphics[width=9cm,angle=-90]{fig4a.ps}
\includegraphics[width=9cm,angle=-90]{fig4b.ps}
\caption{\label{fig:sb}
Surface density profile (top panel) and density profile (bottom panel) at $r=7-9$ kpc for total (dotted lines), thick disk (dot-dashed lines), and thin disk (solid lines) at present.
}
\end{figure}

\begin{figure}
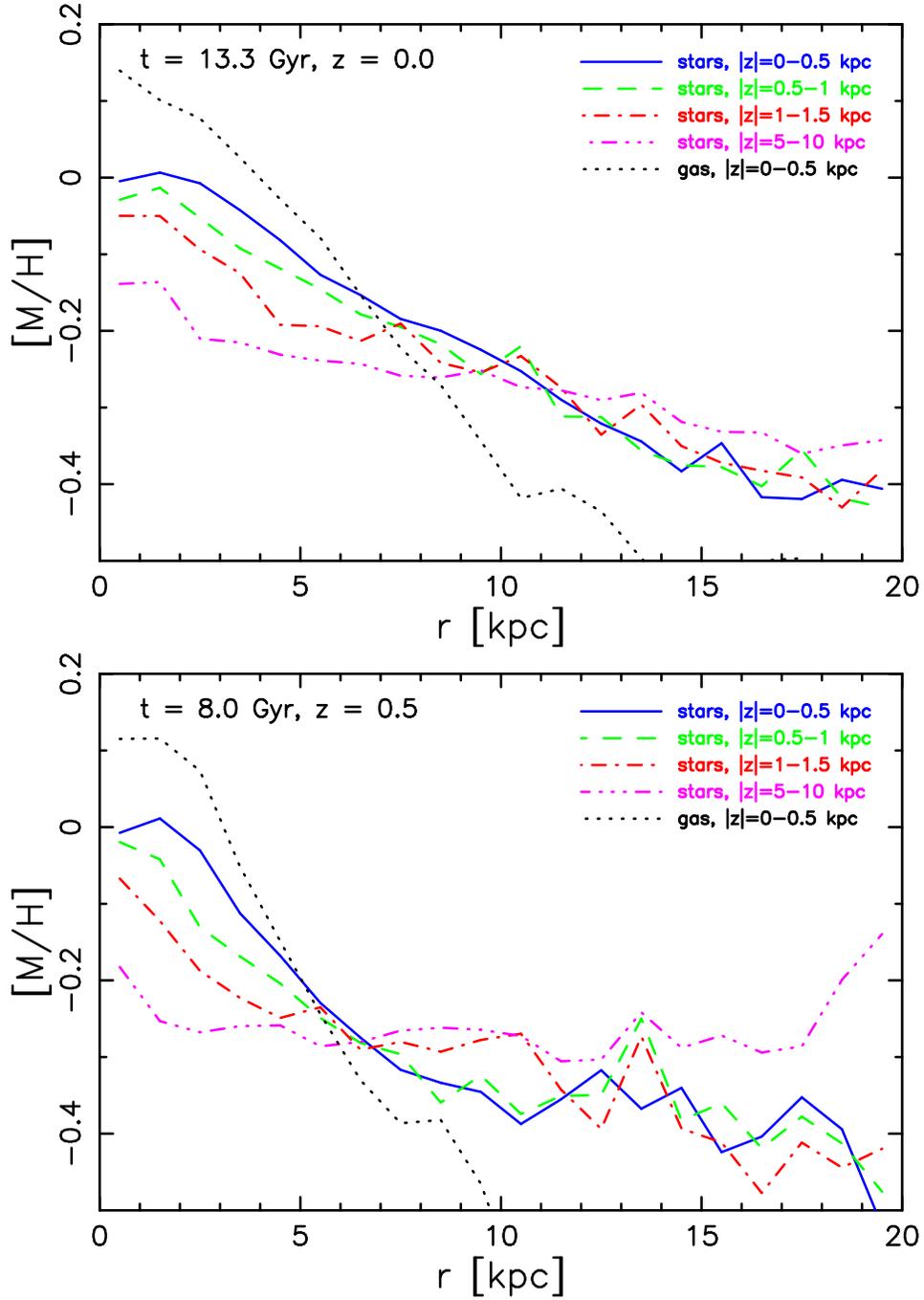

\center 
\includegraphics[width=9cm,angle=-90]{fig5a.ps}
\includegraphics[width=9cm,angle=-90]{fig5b.ps}
\caption{\label{fig:zg}
Radial metallicity gradients of stars at the height $|z|=0-0.5$ (solid line), $0.5-1$ (dashed line), $1-1.5$ (dot-dashed line), and $5-10$ (triple-dot-dashed line) kpc at present (top panel) and at 8 Gyr (bottom panel).
The dotted lines show the radial gradients of gas at $|z|=0-0.5$ kpc.
The metallicities are mass-weighted.
}
\end{figure}

\begin{figure}
\center 
\includegraphics[width=10cm,angle=-90]{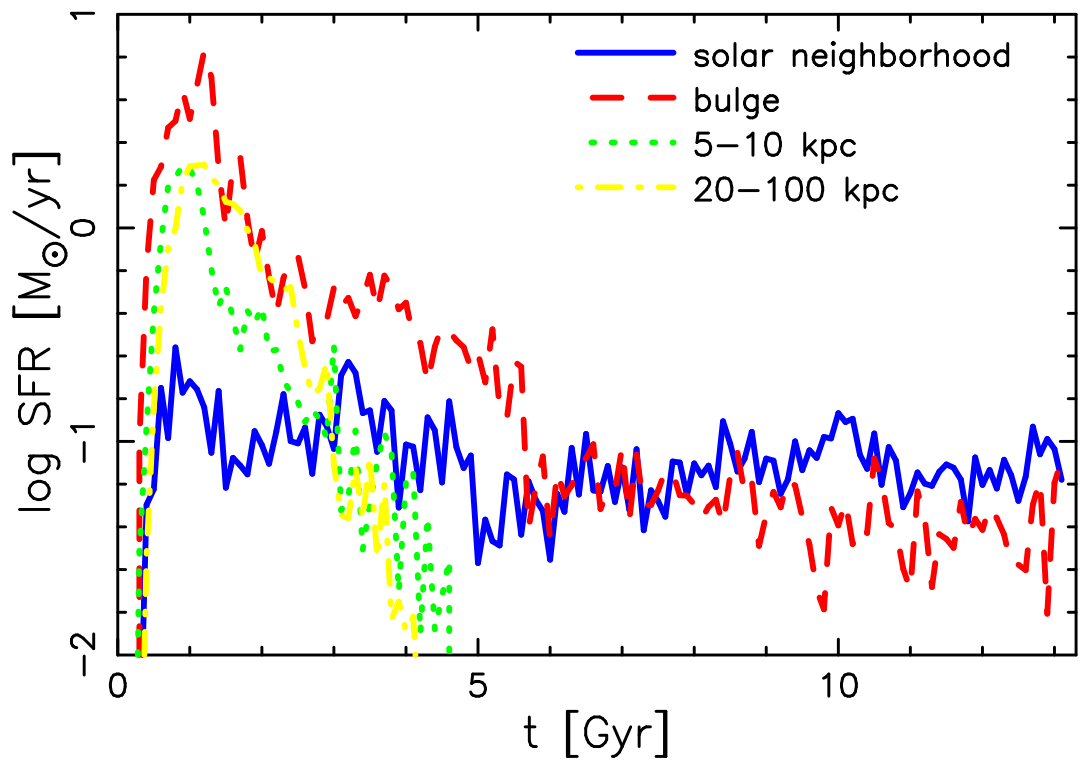}
\caption{\label{fig:sfr}
Star formation histories in the solar neighborhood (solid line), bulge (dashed line), and inner (dotted line) and outer (dot-dashed line) halo at $z=0$.
}
\end{figure}

\begin{figure}
\center 
\includegraphics[width=10cm,angle=-90]{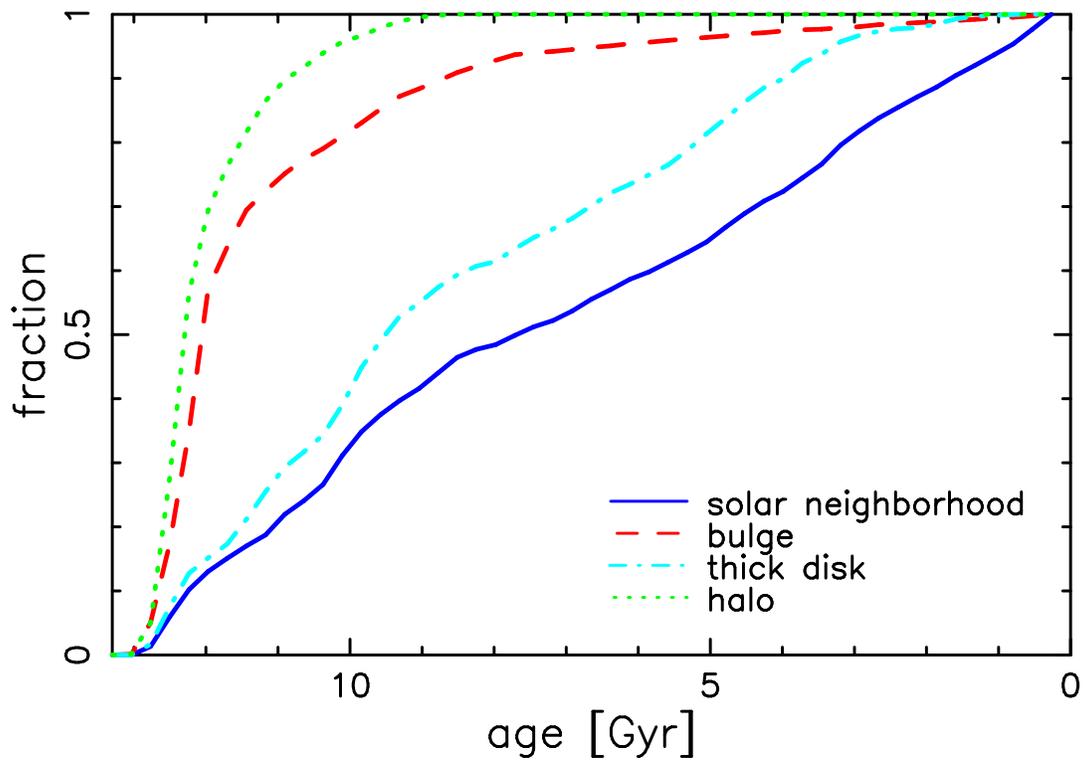}
\caption{\label{fig:agecumnum}
Cumulative functions of the ages of stars in the solar neighborhood (solid line), bulge (dashed line), halo (dotted line), and thick disk (dot-dashed line) at $z=0$.
}
\end{figure}

\begin{figure}
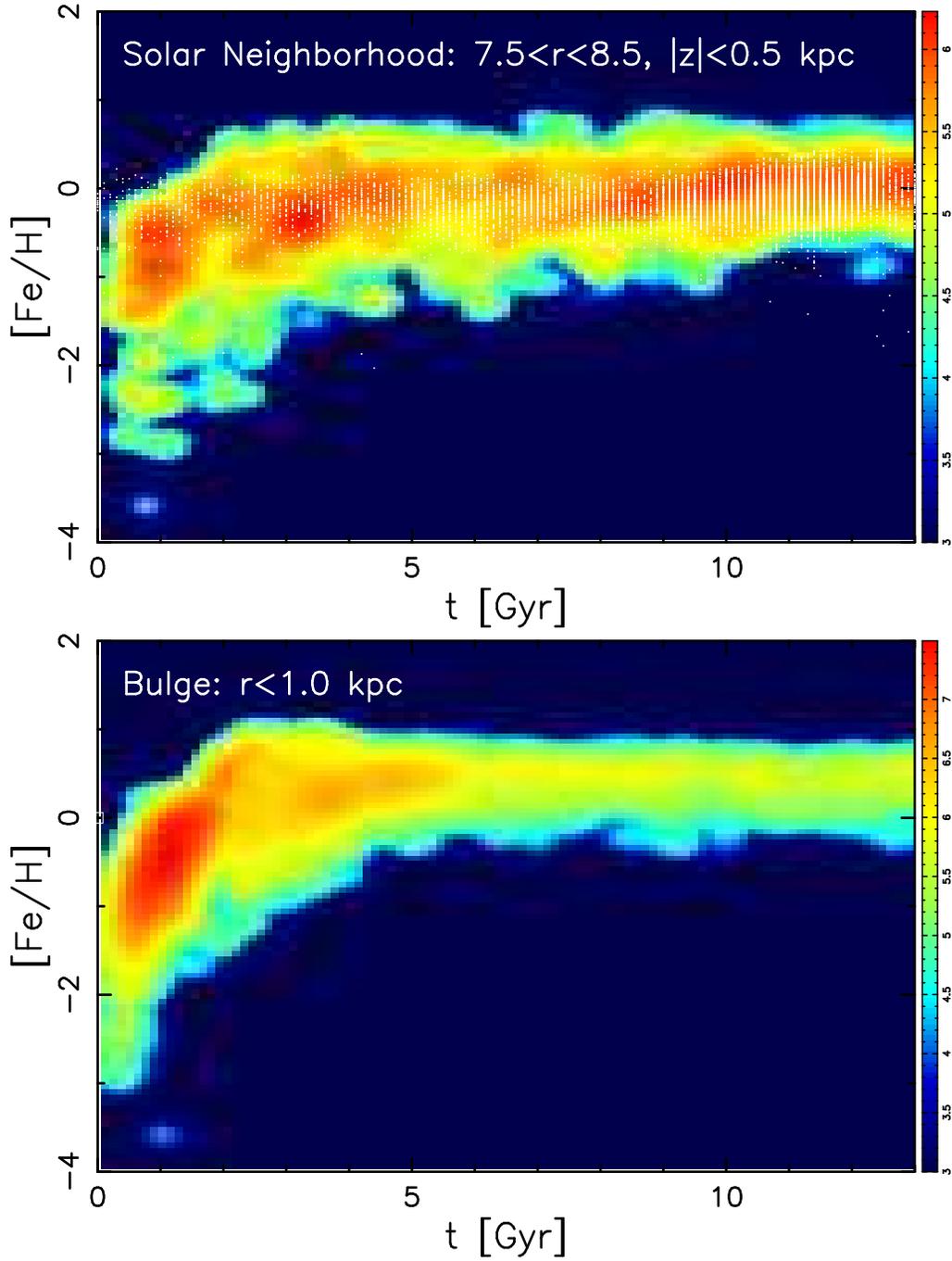

\center 
\includegraphics[width=9cm,angle=-90]{fig8a.ps}
\includegraphics[width=9cm,angle=-90]{fig8b.ps}
\caption{\label{fig:amr}
Age-metallicity relations in the solar neighborhood (upper panel) and bulge (lower panel) at $z=0$.
The contours show the frequency distribution of stars in the simulated galaxy, and red is for the highest frequency.
The white dots show the observations in the solar neighborhood \citep{hol07}.
}
\end{figure}

\begin{figure}
\center 
\includegraphics[width=10cm,angle=-90]{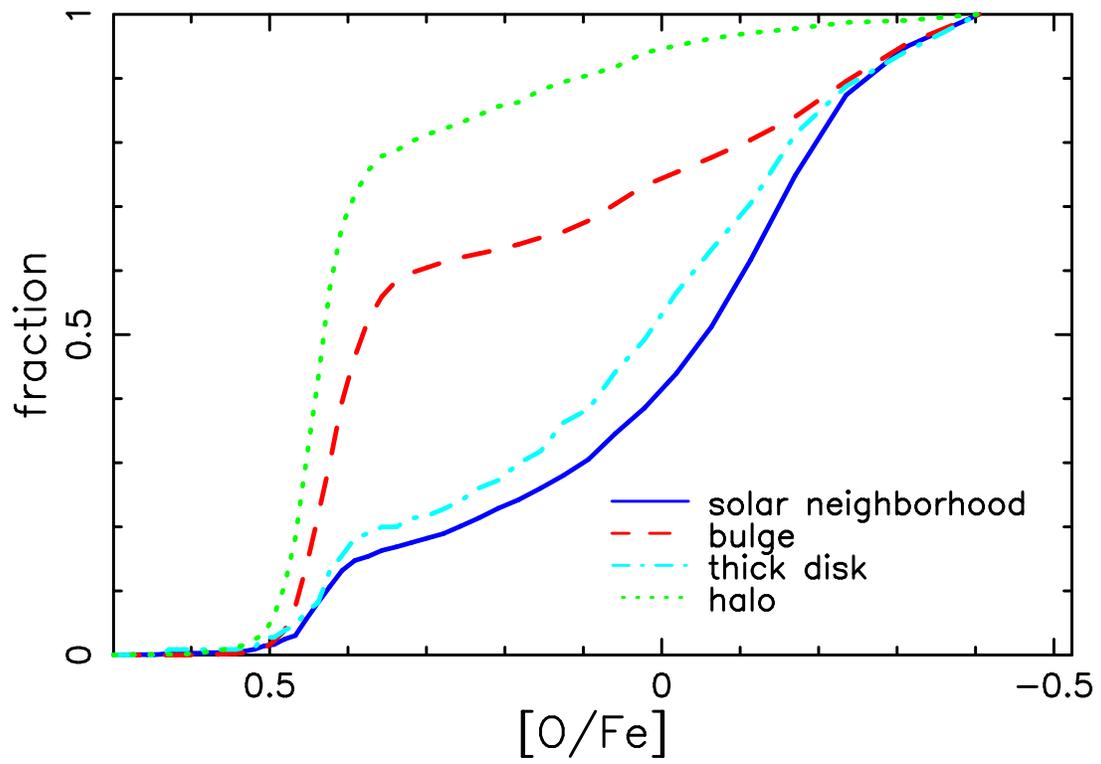}
\caption{\label{fig:afecumnum}
Cumulative functions of [O/Fe] for stars in the solar neighborhood (solid line), bulge (dashed line), halo (dotted line), and thick disk (dot-dashed line) at $z=0$.
}
\end{figure}

\begin{figure}
\center 
\includegraphics[width=10cm,angle=-90]{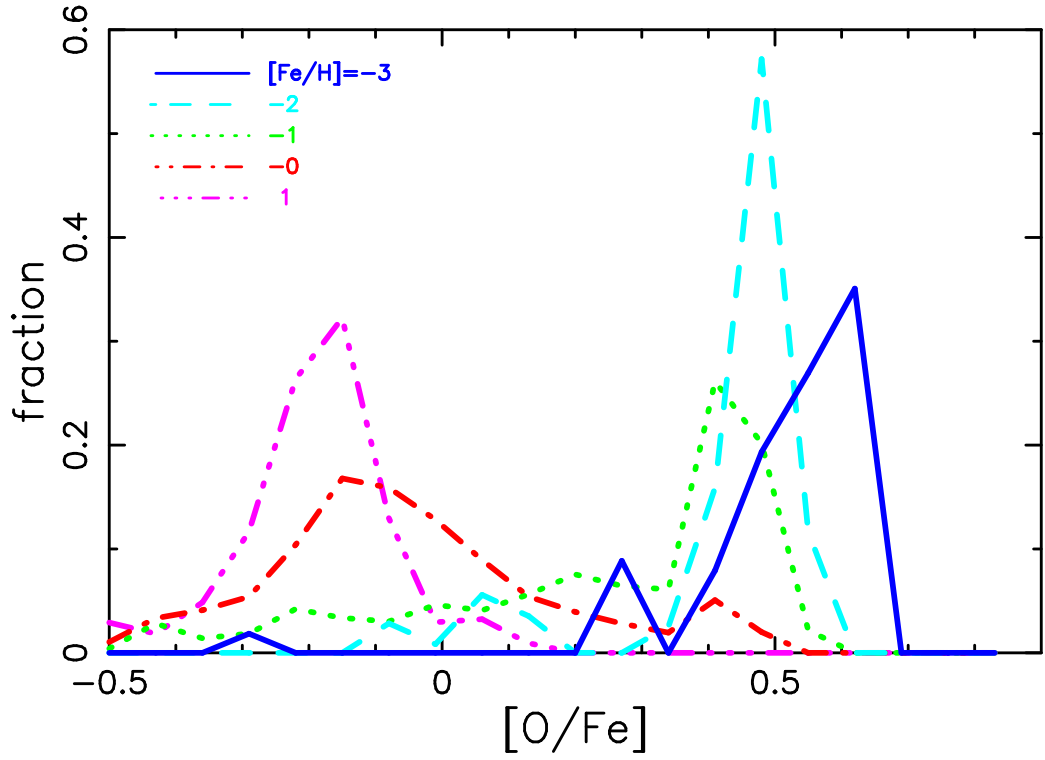}
\caption{\label{fig:afenum2}
Distribution functions of [O/Fe] for stars in the solar neighborhood ($z=0$) at [Fe/H] = $-3.5$ to $-2.5$ (solid line), $-2.5$ to $-1.5$ (dashed line), $-1.5$ to $-0.5$ (dotted line), $-0.5$ to $0.5$ (dot-dashed line), and $0.5$ to $1.5$ (triple-dot-dashed line).
}
\end{figure}

\begin{figure}
\center 
\includegraphics[width=10cm,angle=-90]{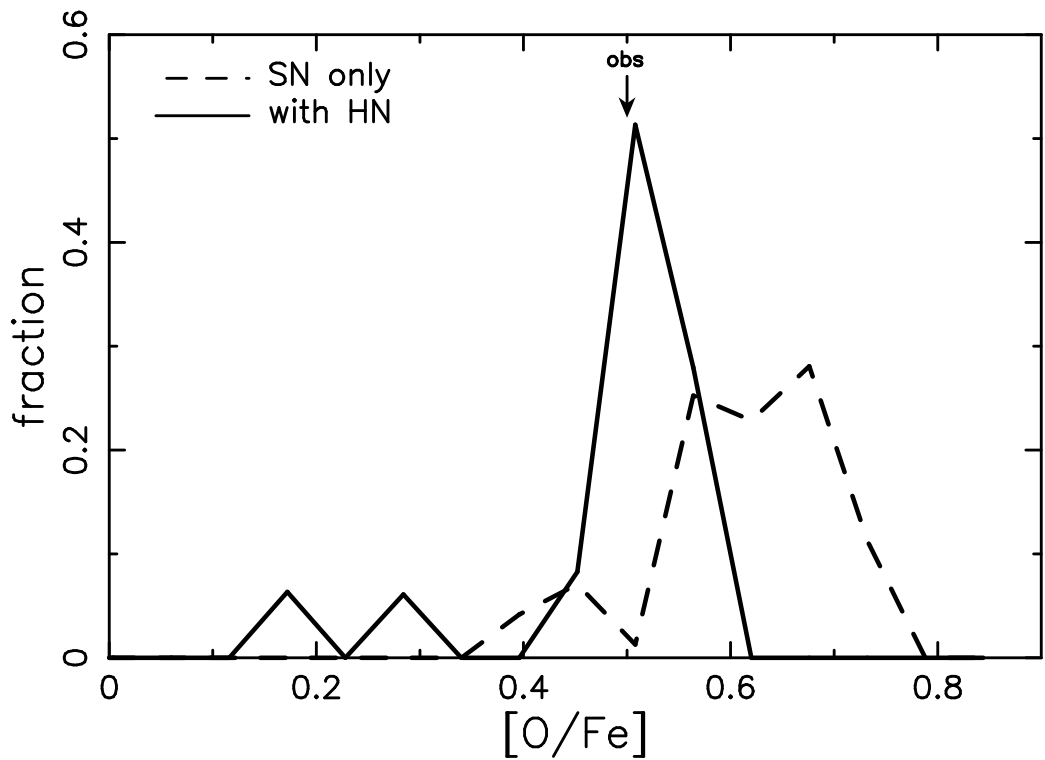}
\caption{\label{fig:afenum19}
Distribution functions of [O/Fe] for stars in the solar neighborhood ($z=0$) at $\le -4.0$ [Fe/H] $\le -1.9$ for the models with (solid line) and without (dashed line) hypernova feedback.
The arrow shows the peak value of observations.
}
\end{figure}

\begin{figure}
\center 
\includegraphics[width=10cm,angle=-90]{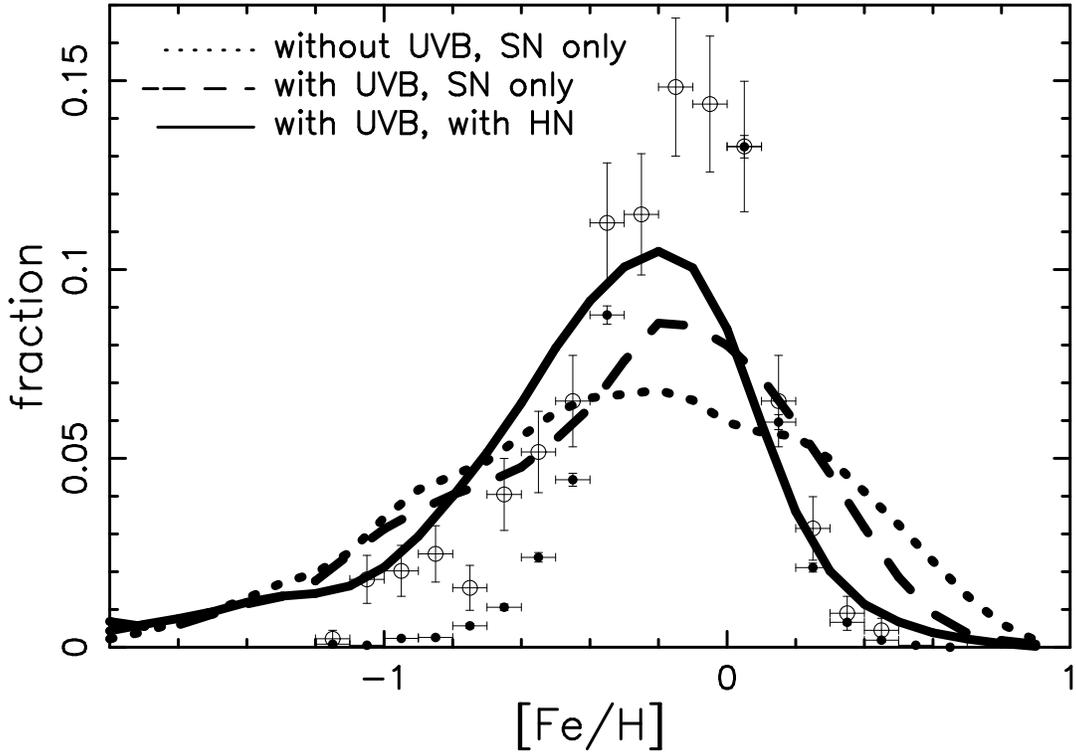}
\caption{\label{fig:mdf}
Metallicity distribution functions of stars in the solar neighborhood at $z=0$ for the models without UV background radiation and hypernovae (dotted line), with UV background radiation but without hypernovae (dashed line), and with UV background radiation and hypernovae (solid line).
The dots with errorbars are for the observations taken from \citet[open circles]{edv93} and \citet[filled circles]{hol07}.
}
\end{figure}

\begin{figure}
\center 
\includegraphics[width=10cm,angle=-90]{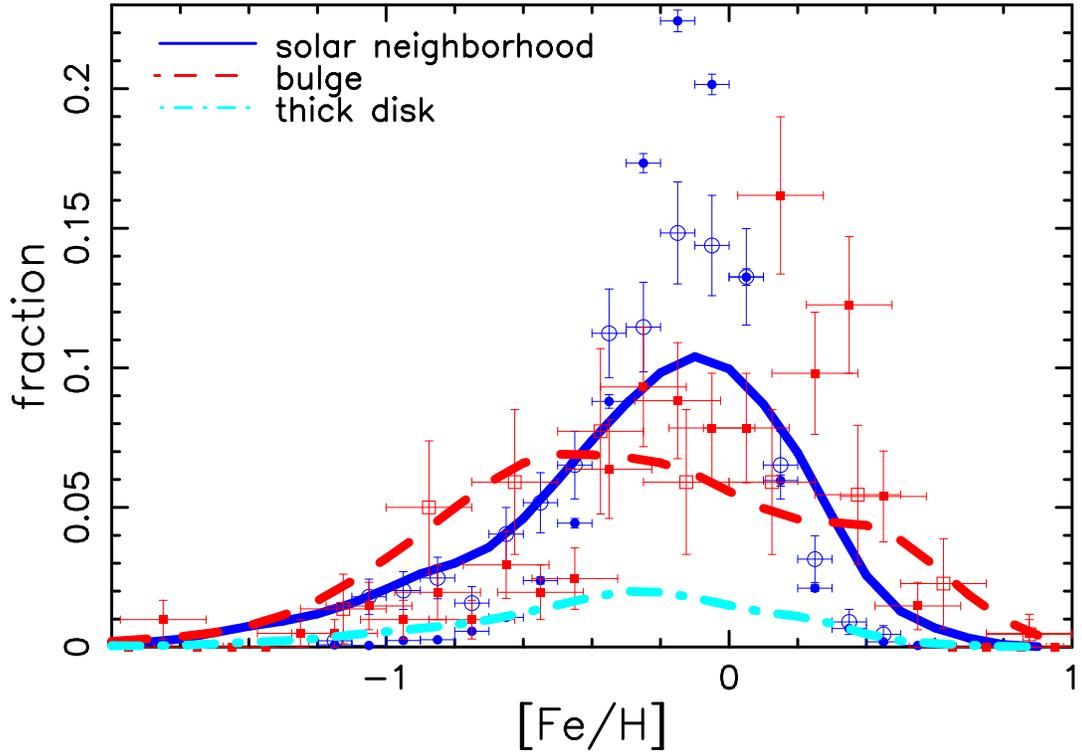}
\caption{\label{fig:mdf2}
Metallicity distribution functions of stars in the solar neighborhood (solid line), bulge (dashed line), and thick disk (dot-dashed line) at $z=0$.
The observational data sources are: \citet{edv93}, open circles; \citet{hol07}, filled circles; \citet{mcw94}, open squares; and \citet{zoc08}, filled squares.
}
\end{figure}

\begin{figure}
\center 
\includegraphics[width=11.5cm]{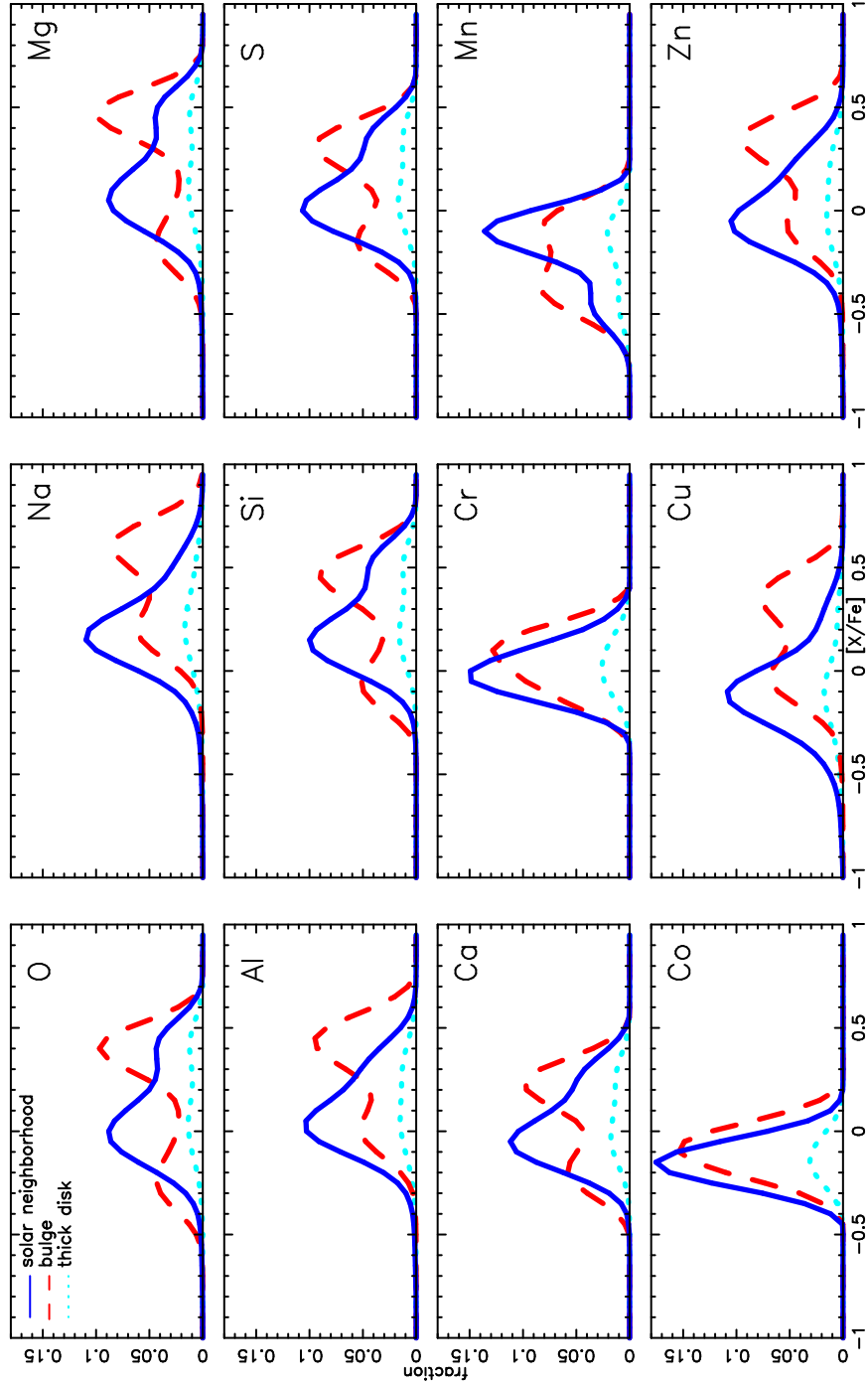}
\caption{\label{fig:xfenum}
Distribution functions of the elemental abundance ratios relative to iron [X/Fe] in the solar neighborhood (solid line), bulge (dashed line), and thick disk (dot-dashed line) at $z=0$.
}
\end{figure}

\begin{figure}
\center 
\includegraphics[width=10cm,angle=-90]{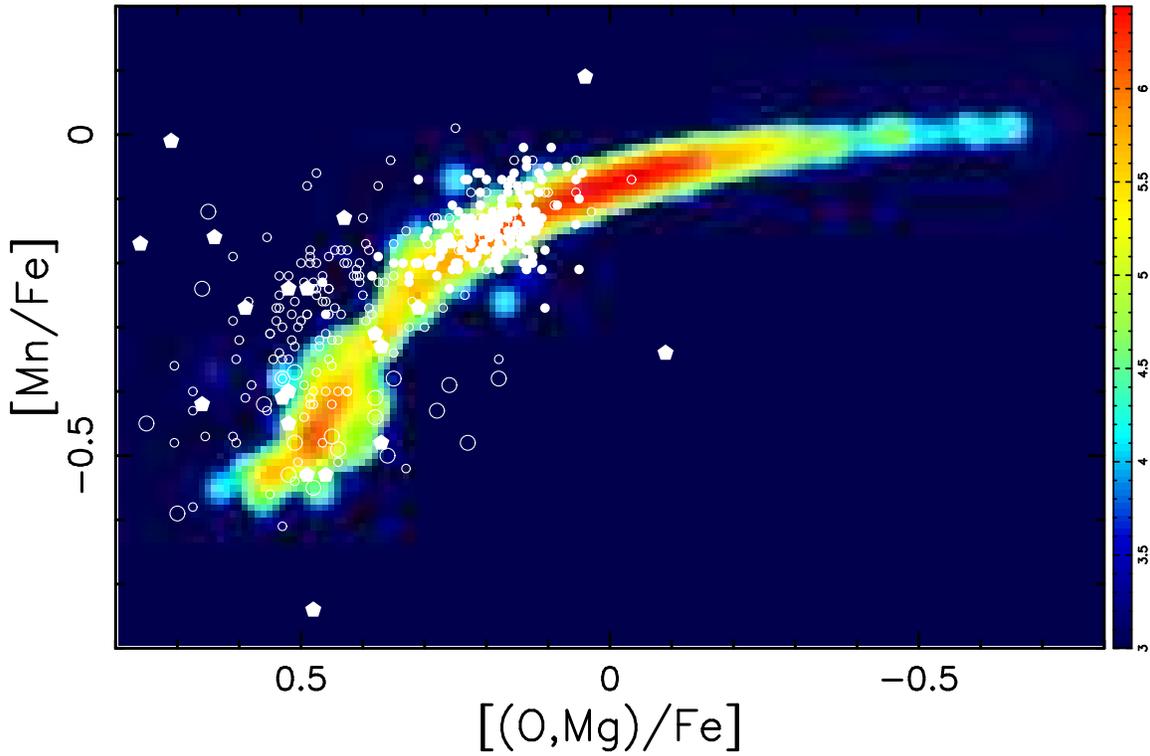}
\caption{\label{fig:mn}
[$\alpha$/Fe]-[Mn/Fe] relation in the solar neighborhood at $z=0$.
The contours show the frequency distribution of stars in the simulated galaxies, where red is for the highest frequency.
The observational data (white dots) are taken from 
\citet{cay04}, large open circles;
\citet{hon04}, filled pentagons;
\citet[2006]{red03} and \citet{red08} for thin (small filled circles) and thick (small open circles) disk stars.
}
\end{figure}

In order to check the resolution and boundary effects, we consider three cases; the total number of particles, the total mass (the half for gas and the rest for dark matter), and the mass resolution for each case are summarized in Table 1.
The initial conditions with the wider region and higher resolution are generated with the same initial perturbation as in the initial condition of the standard model.
Because of star formation, the particle number increases by a factor of $2-3$ at $z=0$. 
The final baryon and gas fractions of the galaxy are also summarized in Table 1 and do not vary very much among these cases.
For the higher resolution run, the calculation is stopped at $z=0.12$.
The gravitational softening length is set to $0.5$ and $0.32$ kpc for the low and high resolutions, respectively.
We obtain qualitatively similar results for all three cases, and consequently, we mainly show the results with the wider initial condition in Figs. \ref{fig:map}-\ref{fig:afenum2} and \ref{fig:mdf2}, and with the standard initial condition in Figs. \ref{fig:afenum05}, \ref{fig:afenum19}, \ref{fig:mdf}, and \ref{fig:xfe}-\ref{fig:mn}.

\section{Results}

\subsection{Star Formation History}

Figure \ref{fig:map} shows the time evolution of the projected density of dark matter, gas, V-band luminosity, and luminosity-weighted stellar metallicity.
After the start of the simulation, the system expands according to the Hubble flow.
The CDM initial fluctuations grow into the structures of nodes and filaments, and small collapsed halos are realized both in dark matter and gas.
In the halos, the gas is allowed to cool radiatively, 
and star formation takes place since $z \sim 15$.
According to the hierarchical clustering of dark halos, subgalaxies merge to form large galaxies, which induces the initial starburst.
Under the CDM picture, any galaxy forms through the successive merging of subgalaxies with various masses. 
In this simulated galaxy, the bulge is formed by the initial starburst that is induced by the assembly of gas-rich sub-galaxies with stellar masses of $\sim 5-10 \times 10^{9} M_\odot$ and gas fractions of $0.2-0.4$ at $z \gtsim 3$.
Because of the angular momentum, the gas accretes onto the plane forming a rotationally supported disk that grows from inside out.
In the disk, star formation takes place with a longer timescale, which is maintained not by the slow gas accretion, but by the self-regulation due to supernova feedback.
Many satellite galaxies successively come in and disrupt, but there is no major merger event after $z\sim2$, which is necessary to retain the disk structure.
Metallicity gradients, increasing toward higher density regions, are generated both in the gas phase and stars from $z \sim 5$ onward.

Figure \ref{fig:map2} shows the B-band (upper panels) and K-band (lower panels) luminosity maps at the present-day for the face-on (left panels) and edge-on views (right panels).
In the B-band, the disk component is visible because of the on-going star formation, while the bulge component is dominant in the K-band.
The galaxy center and the disk plane are determined from the center of gravity and the aliment of the angular momentum vector, respectively.
In the following, $r$ and $z$ denote the radius on the disk plane and the height from the disk plane, respectively.

The bulge has a de Vaucouleurs surface brightness profile with an effective radius of $\sim 1.5$ kpc, and the disk has an exponential profile with a scale length of $\sim 5$ kpc.
In the surface brightness profile, an excess is seen at $r \sim 12$ kpc, which corresponds to the remnant of satellite galaxies.
The bulge in the simulated galaxy seems to be larger than the Galactic bulge, and these properties may be more consistent with M31 \citep[e.g.,][]{kly02}.
Note that it is not feasible to resolve the central blackhole, nuclear bulge, and bar in our simulations.
The total masses inside 1, 10, and 100 kpc are $9.4\times10^9M_\odot$, $2.2\times10^{11}M_\odot$, and $1.1\times10^{12}M_\odot$, respectively.
The baryon fraction is $\sim 0.65$ at $r<2$ kpc, and then decreases to $\sim 0.2$ at $r>8$ kpc.
The rotation velocity of the disk rapidly increases to reach $\sim 230$ km/s at $r \sim 3$ kpc, stays constant until $r \ltsim 20$ kpc, and then gradually decreases until $r\sim40$ kpc.
This is roughly consistent with the observed rotation curve of the Milky Way Galaxy, which shows a plateau at the local standard of rest ($180-250$ km/s) for $3-6 \ltsim r \ltsim 20-60$ kpc \citep{mer92,hon07,lev08}.

In the following, we define the three major components simply from the location of the stars at the present-day: the radius of $7.5 \le r \le 8.5$ kpc and the height of $|z| \le 0.5$ kpc for the solar neighborhood, $r \le 1$ kpc for the bulge, and $5 \le r \le 10$ kpc for the halo.
Thick disk stars are defined from the kinematics using the Toomre diagram in the cylindrical space motion $(V_\Phi, V_\Theta, V_Z)$.
The number of particles significantly drops around the velocity dispersion $\sigma \equiv \sqrt{V_\Phi^2+V_\Theta^2+V_Z^2} \sim 150$ km/s and $\sigma \sim 250$ km/s.
Thus, we define thick disk particles in the range of $\sigma = 160-260$ km/s. This range is different from the observations \citep{ben03,ruc10}, but gives a consistent mass fraction of the thick disk in the simulated galaxy.
With this kinematical definition, the thick disk is smaller and thicker than the thin disk.
The surface density profiles (top panel) and density profiles (bottom panel) at $r=7-9$ kpc for the thin disk (solid lines) and thick disk (dot-dashed lines) are shown in Figure \ref{fig:sb}.
The scale radius is $4.9$ and $4.1$ kpc, the scale height is $1.9$ and $3.2$ for the thin and thick disks, respectively.
These are larger than the 2MASS (Two Micron All Sky Survey) observations \citep{lop02}, which may be due to the lack of resolution.
The mass, mass-to-light ratio, age, metallicity, and abundance ratio of each component are summarized in Table 2.

Figure \ref{fig:zg} shows the radial and vertical metallicity gradients of stars at present (top panel) and at $t=8$ Gyr (bottom panel).
The radial gradient of stars becomes flatter at higher distances from the plane, which seems to be consistent with the SDSS observations \citep{mor10}.
At $|z|=0-0.5$ kpc (solid line), the stellar radial gradient is $\Delta{\rm [M/H]}_*/\Delta r = -0.025$ dex/kpc.
This is flatter than the radial gradient of the ISM (dotted line), which is $\Delta{\rm [M/H]}_{\rm g}/\Delta r = -0.049$ dex/kpc at $|z|=0-0.5$ kpc.
The metallicity radial gradients are steeper at higher redshifts.
At $t=8$ Gyr, $\Delta{\rm [M/H]}_*/\Delta r = -0.062$ dex/kpc and $\Delta{\rm [M/H]}_{\rm g}/\Delta r = -0.094$ dex/kpc for stars and the ISM, respectively.
The time evolution of the ISM gradient is consistent with the observations with planetary nebulae \citep{mac06}.

The resultant star formation histories (Fig. \ref{fig:sfr}) are different for different components.
The initial starburst is induced by the assembly, and most of stars in the bulge (dashed line) and halo (dotted line) are formed at this stage.
In the disk (solid line), the SFR is regulated mainly by supernova feedback.
Figure \ref{fig:agecumnum} shows the cumulative functions of ages of stars in each component in the present-day galaxy.
Most of stars in the present bulge (dashed line) have formed in the first $2$ Gyr, 80\% of bulge stars are older than $10$ Gyr, and 90\% of bulge stars are older than $8$ Gyr.
In the disk (solid line), star formation takes place continuously, and 50\% of solar-neighborhood stars are younger than $8$ Gyr.
In the halo (dotted line), most of stars are as old as bulge, although the numerical resolution of this simulation is not enough to discuss the halo in detail.
However, there does not seem to be very much difference in the SFR between the inner and outer halos.
These star formation histories can be seen for other initial conditions that form disk galaxies at present in our simulations, and also for different initial conditions e.g., those in \citet{sca09} and \citet{san09}.

\subsection{Inhomogeneous Chemical Enrichment}
\label{sec:inhomo}

In chemodynamical simulations, the ISM is not homogeneous at any time.
First of all, there is a local variation in star formation and metal flow by the inflow and outflow of the ISM.
Second, the mixing of heavy elements is caused by supernova feedback.
With our feedback scheme (\S \ref{sec:model}),
star particles obtain heavy elements from the gas particles from which the stars form.
Gas particles obtain heavy elements only when they pass through within a feedback radius of dying star particles.
The amount of heavy elements that the gas particles obtain is physically uncertain, and depend on the modeling of feedback.
Third, the elemental abundance ratios of the metals produced by supernovae depend on the metallicity and mass of progenitors (\S \ref{sec:chem}), i.e., the metallicity and age of star particles.

In our simulations, the metallicity of the first enriched stars reaches [Fe/H] $\sim -3$.
At later times, the star forming region becomes denser, and both metal richer and poorer stars than [Fe/H] $\sim -3$ appear.
Different from one-zone chemical evolution models, the following phenomena occur in the case of inhomogeneous enrichment:
i) The age-metallicity relation is weak.
ii) SNe Ia can affect the elemental abundance ratios at [Fe/H] $\ltsim -1$ even with the metallicity inhibition of SNe Ia.
iii) The scatter of elemental abundance ratios becomes large if the supernova yield depends on progenitor metallicity.
The details are described in the following sections.
As discussed in \S \ref{sec:model}, we run simulations with several feedback parameters and schemes, and show the best results in this paper.

\begin{deluxetable}{lc|ccc|cc|ccc|ccc|cc}
\rotate
\tabletypesize{\scriptsize}
\tablenum{2}
\tablecaption{\label{tab:mw}
Mean Values in the Solar Neighborhood, Bulge, Halo, and Thick Disk in the Simulated Milky Way-type Galaxy:
(1) stellar mass, (2-4) B, V, and K-band mass-to-light ratios, (5-6) mass and luminosity weighted age of stars, (7-8) mass and luminosity weighted metallicity of stars, (9) metallicity of gas, (10-11) mass and luminosity weighted [O/Fe], (12) [O/Fe] of gas, (13) gas fraction, and (14) baryon fraction.}
\footnotesize
\tablewidth{0pt}
\tablehead{
 & (1) & (2) & (3) & (4) & (5) & (6) & (7) & (8) & (9) & (10) & (11) & (12) & (13) & (14) \\
Component & $M_*$ [$M_\odot$] & $M/L_{\rm B}$ & $M/L_{\rm V}$ & $M/L_{\rm K}$ & age [Gyr] & age$_{\rm L}$ [Gyr] & [M/H] & [M/H]$_{\rm L}$ & [M/H]$_{\rm g}$ & [O/Fe] & [O/Fe]$_{\rm L}$ & [O/Fe]$_{\rm g}$ & $f_{\rm g}$ & $b$
}
\startdata
solar neighborhood     & $8.72\times 10^8$ & 1.90 & 2.22 & 0.95 & 7.07 & 3.27 & -0.19 & -0.20 & -0.23 & -0.07 & -0.16 & -0.21 & 0.42 & 0.63\\
bulge      & $4.53\times 10^9$ & 6.92 & 6.15 & 1.51 & 11.17 & 9.03 & -0.01 & -0.03 & 0.14 & 0.01 & -0.10 & -0.32 & 0.22 & 0.62\\
halo       & $1.26\times 10^9$ & 7.92 & 6.28 & 1.69 & 12.02 & 12.07 & -0.27 & -0.38 & -1.63 & 0.24 & 0.25 & -0.85 & 0.02 & 0.06\\
thick disk & $1.78\times 10^8$ & 5.57 & 4.53 & 1.27 & 8.61 & 7.49 & -0.23 & -0.30 & - & -0.04 & -0.08 & - & - & 0.70\\
\enddata
\end{deluxetable}

\subsection{Age-Metallicity Relations}

The chemical enrichment timescale is also different for the different components.
Figure \ref{fig:amr} shows the age-metallicity relations in the present-day galaxy.
In the solar neighborhood (upper panel), the average metallicity reaches [Fe/H] $\sim 0$ at $t \sim 2$ Gyr and does not show strong evolution for $t \gtsim 2$ Gyr.
The scatter in metallicity at given time is caused by the inhomogeneity of chemical enrichment in our chemodynamical model (\S \ref{sec:inhomo}).
As a result, both the average and scatter are in good agreement with the observations (dots) in spite of the uncertainties in the observational estimates of the ages.
At $t \ltsim 5$ Gyr, a week trend is seen in our simulation, which seems to be consistent with the RAVE observations \citep{fre10}.

In the bulge (lower panel), star formation takes place more quickly, and thus the chemical enrichment timescale is much shorter than in the disk. 
The age-metallicity relation shows a more rapid increase than in the disk.
The maximum metallicity reaches super solar ([Fe/H] $\sim 1$) at $t \sim 2$ Gyr.
Although the SFR becomes small after $\sim 5$ Gyr, a few stars form at $\gtsim 5$ Gyr. These have super-solar metallicity in general and the average metallicity does not show time evolution.

\subsection{[$\alpha$/Fe]-[Fe/H] Relations}

The difference in the chemical enrichment timescales results in a difference in the elemental abundance ratios, since different elements are produced by different supernovae with different timescales.
The best known clock is the $\alpha$-elements to iron ratio [$\alpha$/Fe] since SNe Ia produce more iron than $\alpha$ elements with longer timescales than SNe II. Nevertheless it should be noted that low-mass SNe II also provide relatively low [$\alpha$/Fe] because of their smaller envelope mass compared to more massive SNe II.
This mass dependence is important in a system with a low SFR such as dwarf spheroidal galaxies.

Figure \ref{fig:afecumnum} shows the cumulative functions of [O/Fe] of stars in each component in the present-day galaxy.
In the solar neighborhood (solid line), 80\% of stars have [O/Fe] $<0.3$, which indicates a significant contribution from SNe Ia.
In the bulge (dashed line) and halo (dotted line), the chemical enrichment timescale is so short that 60\% and 80\% of stars have [O/Fe] $>0.3$, respectively, although there is a significant number of low [$\alpha$/Fe] stars.
The majority of low [$\alpha$/Fe] stars also have high [Mn/Fe] because of the SN Ia contribution. They are formed at $t \gtsim 1$ Gyr and have non-rotating ($v_\Theta/\sigma \ltsim 0.3$) kinematics.
A small fraction of low [$\alpha$/Fe] stars, however, have low [Mn/Fe] and are formed at $t \ltsim 1$ Gyr. They represent the effect of low-mass SNe II in the inhomogeneous enrichment (the third reason in \S \ref{sec:inhomo}).
Such low [$\alpha$/Fe] stars seem to be observed in SEGUE \citep{bee10}. The fraction of these stars is important for discussions of the formation history of the Galactic halo.
In order to distinguish these two enrichment sources, elemental abundance ratios of iron-peak elements such as Mn are critically important.

\begin{figure}
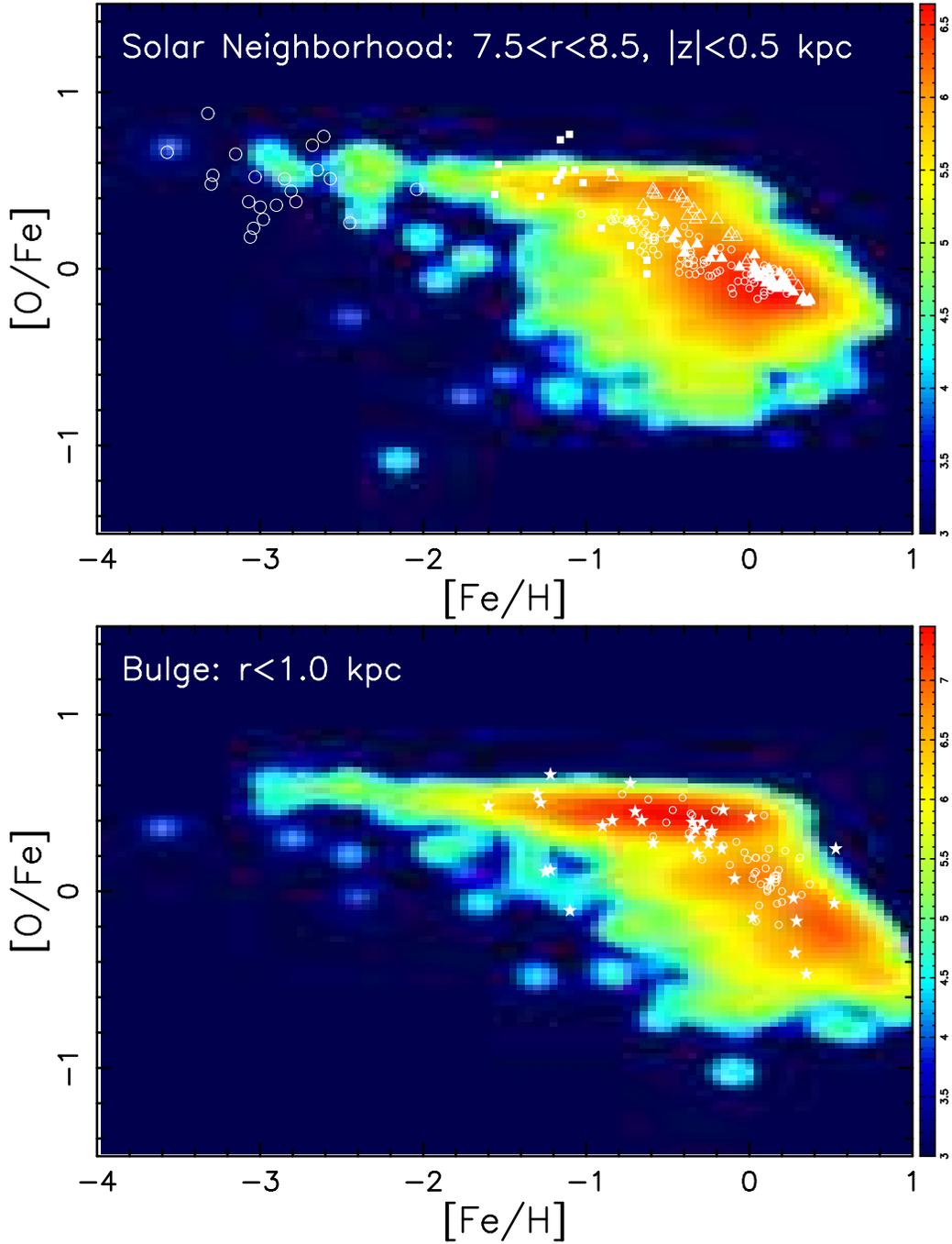

\center 
\includegraphics[width=9cm,angle=-90]{fig10a.ps}
\includegraphics[width=9cm,angle=-90]{fig10b.ps}
\caption{\label{fig:afe}
[O/Fe]-[Fe/H] relations in the solar neighborhood (upper panel) and bulge (lower panel) at $z=0$.
The contours show the frequency distribution of stars in the simulated galaxies, and red is for the highest frequency.
The white dots show the observations of stars:
for the solar neighborhood, 
\citet{edv93}, small open circles;
thin and thick disk stars in \citet{ben04}, filled and open triangles, respectively; 
dissipative component in \citet{gra03}, filled squares;
and \citet{cay04}, large open circles.
For the bulge, \citet{mcw04}, filled stars; and \citet{lec07}, open circles.
}
\end{figure}

Figure \ref{fig:afe} shows the [O/Fe]-[Fe/H] relations, and the other $\alpha$-elements show the same trends.
In the solar neighborhood (upper panel), at the beginning, only SNe II contribute, and [$\alpha$/Fe] shows a plateau ([$\alpha$/Fe]$\sim 0.5$).
Around [Fe/H] $\sim -1$, SNe Ia start to occur, which produce more iron than $\alpha$ elements. This delayed enrichment of SNe Ia causes the decrease in [$\alpha$/Fe] with increasing [Fe/H].
This trend is in great agreement with the observations (dots).
Theoretically, the decreasing point of [$\alpha$/Fe] has been used to put a constraint on the lifetimes of SNe Ia.
\citet{kob98} and \citet{kob09} showed that [$\alpha$/Fe] decreases too early in the double-degenerate scenario and in the single degenerate without metallicity effect since the average lifetime is too short.
The slope of $\Delta$[$\alpha$/Fe]$/\Delta$[Fe/H] depends on the binary parameters in our SN Ia model, but we should note that it is consistent with the observed supernova rates in various types of galaxies (\S \ref{sec:chem}).

A significant scatter is seen in Figure \ref{fig:afe}, and more clearly, the distribution functions of [O/Fe] at given [Fe/H] in the solar neighborhood are shown in Figure \ref{fig:afenum2}.
At [Fe/H] $\ltsim -1$, [O/Fe] is almost constant, and slightly larger for higher metallicity (the peak [O/Fe] $=0.6, 0.5$, and $0.4$ at [Fe/H] $=-3, -2$, and $-1$, respectively) because of the mass dependence of the yields of SNe II.
The scatter of [O/Fe] is caused by the inhomogeneity of chemical enrichment in our chemodynamical model.
At [Fe/H] $\gtsim -1$, the majority of [O/Fe] is lower than $0.2$ and the peak [O/Fe] is $-0.15$.
Around [Fe/H] $\sim -1$, the scatter looks a bit larger than observed (Fig. \ref{fig:afe}).
This may be because the mixing of heavy elements among gas particles is not included in our model (\S \ref{sec:model}).
At [Fe/H] $\ltsim -1$, the scatter is caused from the following two reasons.
First, although SNe Ia cannot occur at [Fe/H] $\le -1.1$ in our SN Ia progenitor scenario, SNe Ia can contribute in the abundances of stars at [Fe/H] $\ltsim -1$.
As discussed in \S \ref{sec:inhomo},
in the inhomogeneous enrichment, it is natural that the enrichment sources have [Fe/H] $> -1.1$ but the stars forming from the ejecta have [Fe/H] $\ltsim -1$ because of the large amount of hydrogen in protostellar clouds. In this case, [Mn/Fe] is as high as $\sim 0$ (see \S \ref{sec:xfe} for more discussion).
Secondly, there is an intrinsic variation in [$\alpha$/Fe] of supernova yields depending on the progenitor mass and metallicity, of which effect is included in our chemodynamical model.
Low-mass SNe II ($10-13 M_\odot$) provide smaller [$\alpha$/Fe] than massive SNe II. In this case, [Mn/Fe] is lower than $0$.
The [$\alpha$/Fe] scatter in the simulation could be larger, since there is also a variation depending on the explosion energy and the remnant mass (neutron star and blackhole) for SNe II and HNe.

From the comparison of the [$\alpha$/Fe] scatter between simulations and observations, it is possible to put constraints on the unsolved physics of supernova explosion and mixing of the ISM.
Figure \ref{fig:afenum05} shows the distribution functions of [O/Fe] for stars in the solar neighborhood at $-0.5 \le$ [Fe/H] $\le -0.3$ for the standard model (solid line) and the models with a longer star formation timescale ($c=0.02$, dashed line), a smaller feedback radius ($r_{\rm FB}=0.5$ kpc, dotted line), and a fixed feedback number ($N_{\rm FB}=50$, dot-dashed line).
The smaller initial condition (Tab. 1) is used here.
The standard model gives much better agreement with the peak value of the observations (arrow) than the other models.

Using stochastic chemical evolution model, \citet{arg02} claimed that the scatter of [$\alpha$/Fe] caused by supernova yields is too large compared to observations and an efficient mixing process is required. 
This is because the nucleosynthesis yields adopted in that paper \citep{tsu95,thi96,nom97a} were different from currently accepted values.
The low [$\alpha$/Fe] values resulted from the larger amount of iron assumed for low-mass SNe II ($0.15 M_\odot$ for $13-15 M_\odot$).
Now the iron mass is estimated from the light curve and spectral fitting, and is $0.07 M_\odot$ for SNe II with $< 20 M_\odot$.
The high [$\alpha$/Fe] values are reduced by including hypernovae.
Figure \ref{fig:afenum19} shows the distribution functions at $-4 \le$ [Fe/H] $\le -1.9$ for the models with (solid line) and without (dashed line) hypernova feedback using the smaller initial condition.
With hypernovae, the scatter around [O/Fe] $\sim 0.5$ is as small as the observations \citep{cay04}.

In the bulge (lower panel of Fig. \ref{fig:afe}), the [$\alpha$/Fe]-[Fe/H] relation is very much different.
The chemical enrichment timescale is so short that the metallicity reaches super solar before SNe Ia contribute.
Thus, the [$\alpha$/Fe] plateau continues to [Fe/H] $\sim +0.3$.
This is roughly consistent with the observations \citep{mcw04,lec07}, except for the O observation in \citet[see \S \ref{sec:xfe} for more discussion]{mcw04}.
In this simulated galaxy, some new stars are still forming in the bulge, which in general have super solar metallicity and low [$\alpha$/Fe] because of the large contribution from SNe Ia.
Although a small fraction of stars with the age of $\sim 3$ Gyr ($t \sim 10.3$ Gyr) do have high [$\alpha$/Fe] ([O/Fe] $=0.3$), which is caused by the inhomogeneous enrichment, stars younger than $1$ Gyr have $-0.5 \le$ [O/Fe] $\le 0$ and $0 \le$ [Fe/H] $\le 0.8$.

\citet{cun07}, however, showed several young metal-rich stars with high [$\alpha$/Fe] in the Galactic center, which suggests additional physical processes such as
(i) local enrichment by the stellar winds of dying stars or SNe II,
(ii) intrinsic variation of supernovae depending on energy and mixing,
or (iii) the selective mass-loss of the gas enriched by SNe Ia.
The bulge wind might be driven by SNe Ia, and most of Fe enriched gas might be selectively blown away before these young stars are formed.
If the observed scatter in [O/H], [O/Fe], and [Ca/Fe] are real, then the scenarios (i) and (ii) may be preferable.

\subsection{Metallicity Distribution Functions}

The metallicity distribution function (MDF) of G-dwarf stars is a stringent constraint on chemodynamical models.
Figure \ref{fig:mdf} shows the present MDF for three different models with the smaller initial condition.
The model with the feedback from the UV background radiation (UVB) and hypernova (solid line) gives much better agreement with the observations in the solar neighborhood.
However, the number of metal-poor stars is still larger than observed, in other words, there is the G-dwarf problem.
In one-zone chemical evolution models, this problem is easily solved by assuming a long timescale of gas infall \citep[e.g.,][]{tin80}, but we cannot control the infall by hand in chemodynamical simulations.
The mass accretion timescale is mainly determined from the hierarchical clustering of dark halos.
Although hypernova feedback could help in generating outflow, the mass accretion history is not affected very much.
In our models, 77\% of dark matter (45\% of total mass) is accreted inside 10 kpc in the first 1 Gyr, then the rest is gradually accreted.
There is a room to change the star formation and chemical enrichment timescales.
With the UVB, the initial starburst is suppressed by a factor of $1.5$, which results in the longer duration of star formation.
With HNe, the SFR is a little suppressed at $t \gtsim 3$ Gyr, and the metal ejection is significantly changed.

Figure \ref{fig:mdf2} shows the present MDF in the different components of the simulated galaxy for the best model with the wider initial condition.
In the solar neighborhood (solid line), the MDF is roughly consistent with the previous observations \citep{edv93,wys95}.
However, the Geneva-Copenhagen survey showed a narrower MDF (\citealt{hol07}, updated from \citealt{nor04}).
If this is valid, the G-dwarf problem becomes more serious.
In the bulge (dashed line), \citet{mcw94} and \citet{sad96} showed a broad MDF, from which \citet{nak03} discussed the metal-rich population in the bulge.
Also in our simulations, the metal-rich peak at [Fe/H] $\sim 0.4$ is caused by the delayed enrichment of SNe Ia.
However, \citet{zoc08} showed a narrower MDF with a large spectroscopic sample.
There seems to be the G-dwarf problem also for the Galactic bulge.
The lack of metal-rich tail in the MDF may suggest that the star formation is truncated in the bulge, possibly by an SNe Ia driven bulge wind.

The other solutions of the G-dwarf problem are, as summarized in \citet{tin80}, pre-enrichment and/or variable IMF.
If the gas is enriched by the first stars before it accretes onto the Galaxy, the number of metal-poor stars can be small.
A bulge wind driven by SNe Ia could also work as pre-enrichment in the disk.
If the IMF at low-metallicity is top-heavy as suggested by the simulations of primordial star formation \citep{bro04}, larger amounts of metals and heating sources are provided.
If no low-mass stars can be formed at low-metallicity, the number of metal-poor G-dwarf stars obviously becomes zero.
We should note, however, that the observed abundance patterns of the extremely metal-poor (EMP) stars rule out large contributions from pair instability supernovae that result from $\sim 140-270M_\odot$ stars.
Stars with $\ltsim 100 M_\odot$ and $\gtsim 300 M_\odot$ are usually supposed to collapse to blackholes, but may be able to explode as core-collapse supernovae, which do not conflict with the observations \citep{ohk06}.
Or, simply, simulations with higher resolution may solve the G-dwarf problem.

\subsection{[X/Fe]-[Fe/H] Diagrams}
\label{sec:xfe}

\begin{figure}
\center 
\includegraphics[width=11.5cm]{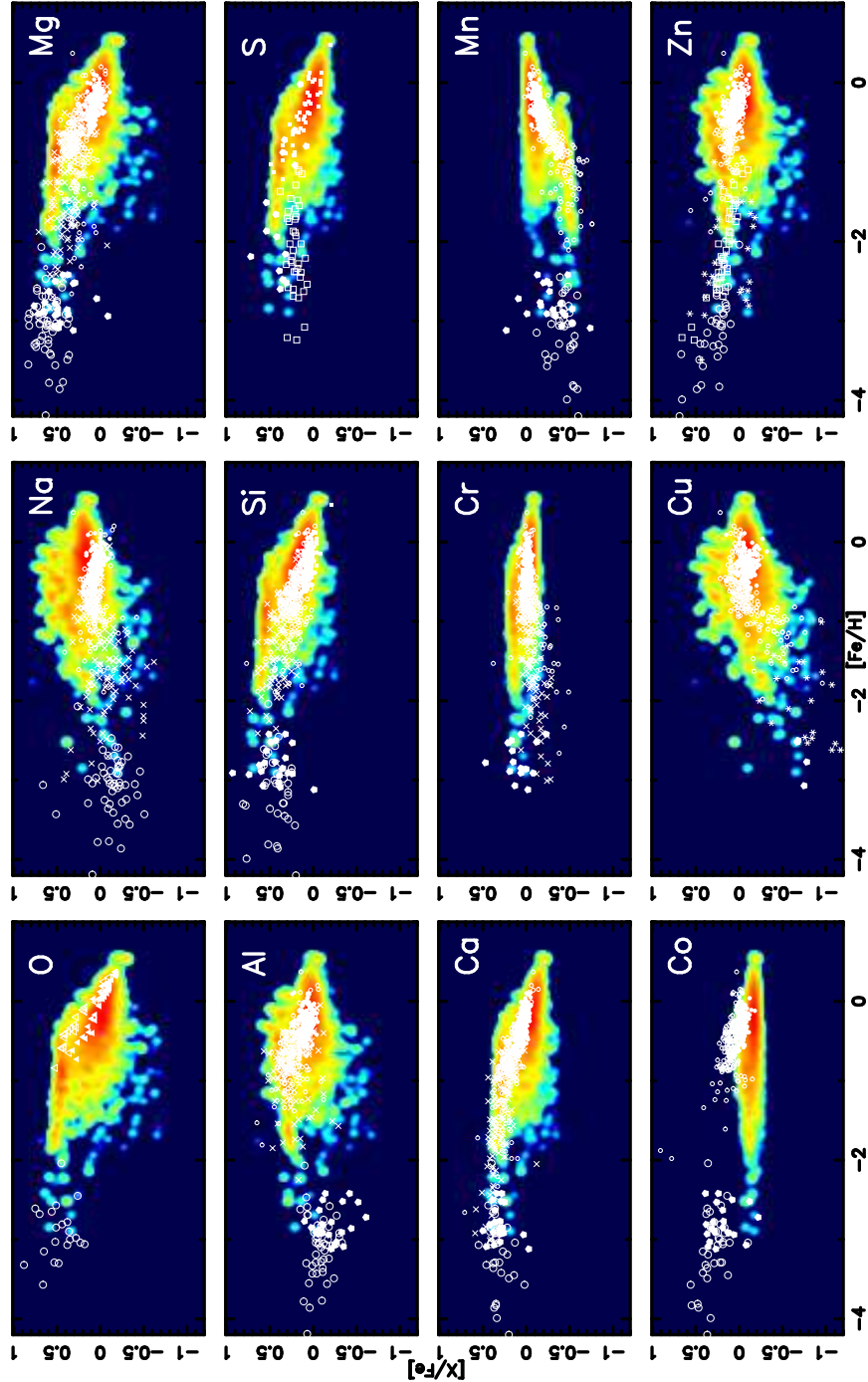}
\caption{\label{fig:xfe}
[X/Fe]-[Fe/H] relations in the solar neighborhood at $z=0$.
The contours show the frequency distribution of stars in the simulated galaxies, where red is for the highest frequency.
The observational data (white dots) are taken from
\citet{cay04}, large open circles;
\citet{hon04}, filled pentagons;
\citet{ful00}, crosses;
\citet[2006]{red03} and \citet{red08} for thin (small filled circles) and thick (small open circles) disk stars.
For O, \citet{ben04} for thin (filled triangles) and thick (open triangles) disk stars.
For Si, S, and Zn, \citet{che02}, filled squares; \citet{tak05}, filled pentagon; \citet{nis07}, open squares.
For Cu and Zn, \citet{pri00}, asterisks.
}
\end{figure}

\begin{figure}
\center 
\includegraphics[width=11.5cm]{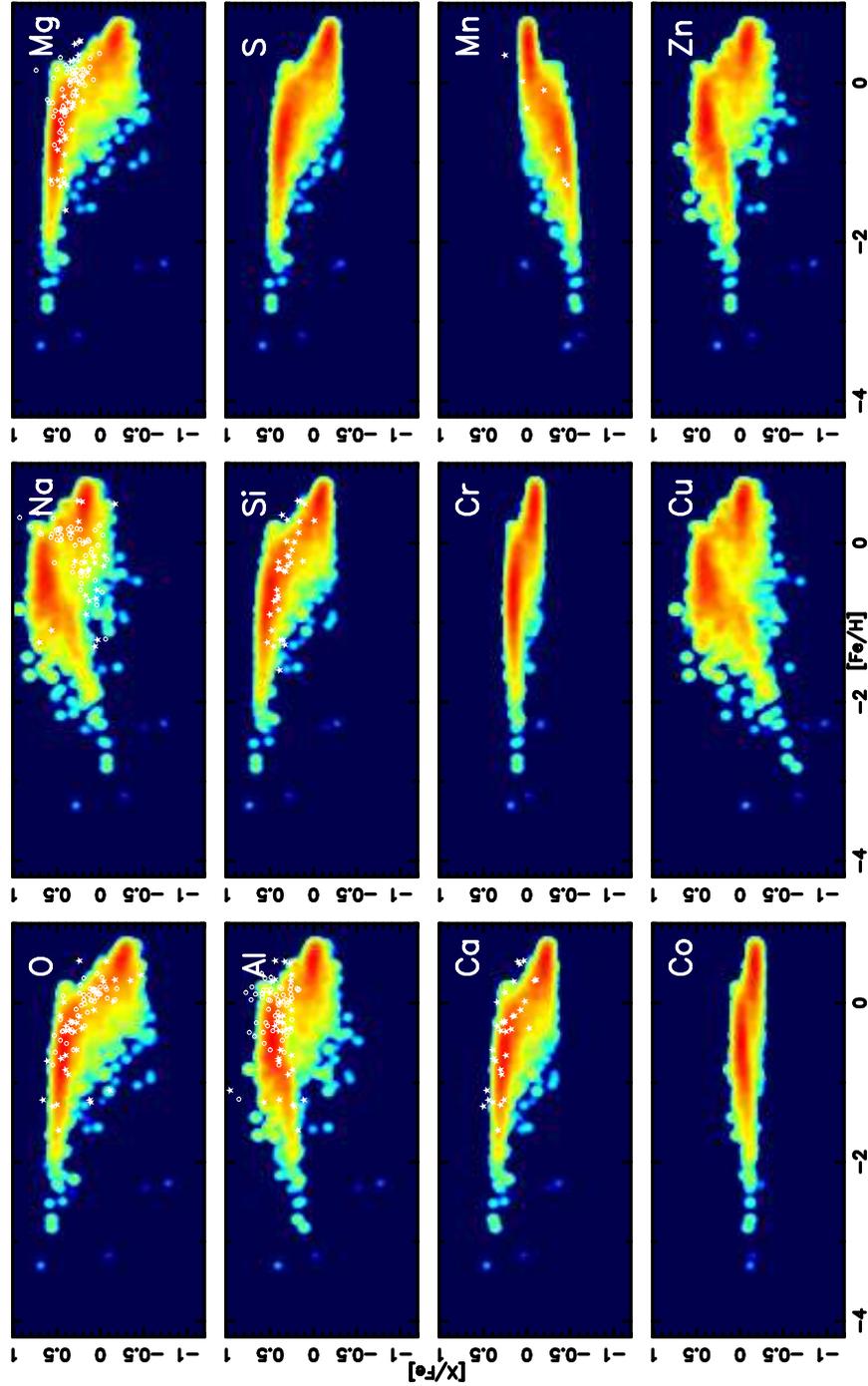}
\caption{\label{fig:xfe-bulge}
Same as Fig. \ref{fig:xfe} but for the bulge.
The observational data (white dots) are taken from
\citet{mcw04}, filled stars;
and \citet{lec07}, open circles.
}
\end{figure}

Finally, using our chemodynamical simulation, we predict the frequency distributions of the elements from O to Zn as a function of time and location.
Figures \ref{fig:xfe}-\ref{fig:xfe-thick} show the mass density of stars in the [X/Fe]-[Fe/H] diagrams at present.
These results can be statistically compared with the next generation of observations from high resolution and multi-object spectrographs such as HERMES. These instruments enable us not only to study the formation and evolution history of the Galaxy but also to update stellar evolution and supernova physics.

Figure \ref{fig:xfe} is for the solar neighborhood.
The dots are for the observations of individual stars, and they almost perfectly overlap with the contours of our simulation.
In order to minimize the systematic error among various methods of observational data analysis, we consider only data from \citet[hereafter R03]{red03}, \citet{red06}, \citet{red08} at [Fe/H] $\ltsim -1$, \citet{ful00} at $-2 \ltsim$ [Fe/H] $\ltsim -1$, and \citet[hereafter C04]{cay04} and \citet[hereafter H04]{hon04} at [Fe/H] $\ltsim -2$.
As in K06, the 3D corrections of $-0.24$ for O in C04, and the NLTE corrections of $+0.1$ and $+0.5$, respectively, for Mg and Al in H04 are applied.
For Mg, Na, Al, and K, C04's data are updated by \citet[2008, 2010, hereafter A10]{and07}  with the NLTE corrections.
For O, S, Cu, and Zn, we employ other data sources as noted below.

\begin{itemize}
\item {$\alpha$ elements} ---
O, Mg, Si, S, and Ca show the same relation; a plateau at [Fe/H] $\ltsim -1$, and then a decrease due to the delayed enrichment of SNe Ia (see KN09 for the SN Ia progenitor models).
For O and S, an increasing trend toward lower [Fe/H] reported by \citet{isr98} and \citet{isr01}, respectively, is affected by the NLTE and 3D effects, and is not seen in our simulation.
For O, there is an offset between the lines used in the abundance analysis. To compare with the observations at low metallicity, we plot \citet{ben04}'s data with the forbidden line [OI] 6300{\AA} since an empirical relation using [OI] and OI triplet at 7775{\AA} is used in R03.
For S, we plot data from \citet{tak05} and \citet{nis07} at low metallicity, and \citet{che02} at high metallicity, where more lines are used than in R03. In R03, the scatter of [S/Fe] is large and the decreasing trend toward low metallicity is not clearly seen.
At [Fe/H] $\gtsim -1$, [Si/Fe] may be slightly larger, and [Ca/Fe] may be slightly smaller than the observations, which are due to the SN Ia yields.

\item {Odd-Z elements} ---
Na, Al, and Cu show a decreasing trend toward lower metallicity since the nucleosynthesis yields of these elements strongly depend on the metallicity of progenitor stars (see K06 for the details).
Na and Al also show a decreasing trend toward higher metallicity due to SNe Ia, which is shallower than that of $\alpha$ elements.
[Na/Fe] at [Fe/H] $\gtsim -1$ may be slightly larger than observed.
The large scatter is caused by the metallicity dependence of yields.
As discussed in \S \ref{sec:inhomo}, in our chemodynamical simulations, there is only a week correlation between the metallicities of supernova progenitors and those of the surrounding ISM.
If the ejected metals are diluted, the ISM metallicity and the metallicity of stars that form from the ISM become lower than the progenitor metallicity.
The scatter starts increasing at [Fe/H] $\sim -2$ in this simulation, which looks consistent with the observations. Observations of an unbiased sample with the NLTE analysis will put a constraint on the mixing process in the chemodynamical simulations.

\item {Iron-peak elements} ---
The scatter of Cr and Co is very small since there is not very much variety in the SN II yields and [(Cr, Co)/Fe] is almost $\sim 0$ for SNe Ia.
For Cr, since a dependence on the temperature of the observed stars has been reported, we plot the Cr II abundance from H04 for low metallicity, which is consistent with our simulation.
For Co, an increasing trend toward lower metallicity is seen in the observations, which is not realized in the simulation.
This is due to the input of supernova yields, and cannot be solved with different star formation histories or initial conditions.
This could be solved if we adopt higher energy and/or higher fraction of HNe where more Co and Zn are synthesized.

\item {Zinc} ---
Zn is one of the most important elements for supernova physics.
[Zn/Fe] is about $\sim 0$ for a wide range of metallicities, which can only be generated by such large contribution of HNe.
In detail, there is a small oscillating trend; [Zn/Fe] is $0$ at [Fe/H] $\sim 0$, increases to be $0.2$ at [Fe/H] $\sim -0.5$, decreases to be $0$ at [Fe/H] $\sim -2$, then increases toward lower metallicity.
This is characteristic in our SN Ia model (KN09), and is consistent with the observations as found by \citet{sai09}.
Theoretically, Zn production depends on many parameters;
$^{64}$Zn is synthesized in the deepest region of HNe, while neutron-rich isotopes of zinc $^{66-70}$Zn are produced by neutron-capture processes, which are larger for higher metallicity massive SNe II.
The scatter at $-1.5 \ltsim$ [Fe/H] $\ltsim -0.5$ is larger than observed, which is caused by the dependence of the Zn yield on the metallicity and mass.
The increasing trend toward lower metallicity in the observations \citep{pri00,nis07} could be generated with the same effects as for Co.

\item {Manganese} ---
Mn is a characteristic element and is more produced by SNe Ia relative to iron.
From [Fe/H] $\sim -1$, [Mn/Fe] shows an increasing trend toward higher metallicity, which is caused by the delayed enrichment of SNe Ia.
Mn is also an odd-Z element, and the Mn yield depends on the metallicity both for SNe II and SNe Ia.
\citet{fel07} showed a bit steeper slope at [Fe/H] $>0$, which would be generated by the metallicity dependence of SN Ia yields.
At [Fe/H] $\ltsim -2$, the scatter in the simulation is a bit smaller than observed.
Since SNe Ia do not contribute at such low metallicity, the [Mn/Fe] scatter, if it is real, should stem from the SN II yields depending on the remnant mass and the electron fraction $Y_{\rm e}$ that is related to the mechanism of supernova explosion.
In the supernova ejecta, Mn and Cr are synthesized in the relatively outer regions compared to Cr, Zn and the majority of Fe.
Therefore, relatively high [(Mn, Cr)/Fe] and low [(Co, Zn)/Fe] are obtained with faint SNe with relatively large remnant masses (blackholes; \citealt{kob10b}).
\end{itemize}

In our figures, we take the solar abundance from \citet{and89}, which is different for some elements from the solar abundance updated by \citet{asp10}.
This means the normalization should be changed both for observations and simulations.
When the observational data are analyzed with 3D and NLTE effects, the agreement in Figures \ref{fig:xfe}-\ref{fig:xfe-thick} may be or may not be modified.
For example, one of the NLTE models shows [Mn/Fe] $\sim 0$, [Cr/Fe] $\sim 0$, and steeper increase in [Co/Fe] toward lower metallicity (\citealt{ber08}, 2010, \citealt{ber10}). 

Figure \ref{fig:xfe-bulge} is for the bulge, where the observational data are taken from \citet{mcw04} and \citet{zoc08}.
As shown in Fig. \ref{fig:afe}, the $\alpha$-element plateau continues to [Fe/H] $\sim 0.3$, then the second component with a mild decreasing trend are generated at $-0.5\ltsim$ [$\alpha$/Fe] $\ltsim 0$ from the SN Ia contribution.
Very similar trends are seen for O, Mg, Si, S, and Ca.
This is roughly consistent with the observations, although the double peak distributions are not seen in the observations.
For O, \citet{mcw04} showed a steeper decrease than Mg at [Fe/H] $\gtsim 0$.
Theoretically, it is not easy to produce such a different trend of O from Mg since these two elements are synthesized in the similar region and ejected to the ISM in a similar way.
The contribution from AGB stars is negligible.
Such a decrease in O/Mg might require some additional effects that are not included in our stellar evolution models such as strong stellar winds or a process that causes the change in the C/O ratio.
For other $\alpha$ elements, explosion energy and remnant mass may cause a variation in (O, Mg)/(Si, S) and (O, Mg)/Ca, respectively.
These should be discussed with future observations with a larger sample and with nucleosynthesis yields that cover the parameter spaces.
The separation between the two populations also depends on the IMF, and the Salpeter IMF is adopted in this simulation. The IMF will be constrained from the elemental abundance ratios.

The odd-Z elements and also Zn show the increasing component from [Fe/H] $\sim -3$ to $\sim 0$, and the second flat component without any trend at [(Na, Al, Cu, Zn)/Fe] $\sim 0$ at $0 \ltsim$ [Fe/H] $\ltsim 1$.
Different from Fig. \ref{fig:xfe}, the metallicity dependence of Na, Al, and Cu yields are not seen very much in the bulge since the chemical enrichment timescale is short.
[Mn/Fe] shows the increasing component from [Fe/H] $\sim -2$ to $\sim 0$, which continues to the second flat component at [Mn/Fe] $\sim 0$ at $0 \ltsim$ [Fe/H] $\ltsim 1$. This is also due to the smaller contribution from SNe Ia in the bulge.
The observations also show the high [Na/Fe] and [Al/Fe] at [Fe/H] $\gtsim -1$, and the increasing trend of [Mn/Fe], although the double peak distributions are not visible.

Figure \ref{fig:xfe-thick} is for the thick disk, where the observational data are taken from \citet{pro00}, \citet{ben04}, \citet{red06}, and \citet{red08}.
When we select the thick disk stars from kinematics in the solar neighborhood, the [$\alpha$/Fe] plateau is dominant, which is similar to the bulge.
This is also consistent with the RAVE observations \citep{ruc10}.
We should note, however, that the split trends observed in the solar neighborhood for the [$\alpha$/Fe] trends associated with the thick and thin disks are not clearly seen in the predicted distributions.
The formation timescale of the thick disk is $3-4$ Gyr in our simulation, which is shorter than the thin disk but longer than the bulge.
In Figures \ref{fig:agecumnum}, \ref{fig:afecumnum}, and \ref{fig:mdf2}, the functions of the thick disk are also shown (dot-dashed lines).
For thick disks stars, the age is older and [O/Fe] is higher than thin disk stars, and the peak metallicity is 0.2 dex lower than the thin disk.

Figure \ref{fig:xfenum} shows the distribution functions of [X/Fe] in the solar neighborhood (solid line), bulge (dashed line), and thick disk (dotted line).
$\alpha$ elements show the double peak distributions, and the [$\alpha$/Fe]-rich peak is at the same [$\alpha$/Fe] for the solar neighborhood, bulge, and thick disk.
On the other hand, Na, Al, Cu, and Zn show the single peak distributions in the solar neighborhood, but the double peak distributions in the bulge.
On the average, Na, Al, Cu, and Zn are much more enhanced in the bulge, and (Na, Al, Cu, Zn)/$\alpha$ is larger than in the solar neighborhood.
In the thick disk, the [(Na, Al, Cu, Zn)/Fe] distributions are rather similar to the solar neighborhood because the metallicity is not as high as in the bulge.
The stellar population of the thick disk is neither thin-disk like nor bulge like (see Tab. \ref{tab:summary}).
For the thick disk stars, the star formation timescale is as short as in the bulge, but the chemical enrichment efficiency is not as high as in the bulge.
This is because half of the thick disk stars have already formed in satellite galaxies before they accrete onto the disk, and the metals have been ejected from the satellite galaxies by the galactic winds.

\subsection{[$\alpha$/Fe]-[Mn/Fe] Diagram}

In Figure \ref{fig:mn}, for the solar neighborhood of the present-day galaxy, [Mn/Fe] is plotted against [$\alpha$/Fe]=([O/Fe]+[Mg/Fe])/2, which clearly shows the sequence of the SN Ia contribution.
With SNe II and HNe only, [$\alpha$/Fe] is as high as $\sim 0.5$, and [Mn/Fe] is as low as $\sim -0.5$. With more SNe Ia, [$\alpha$/Fe] decreases, while [Mn/Fe] increases.
The three populations of the observed stars are along this trend;
i) the EMP stars (O in C04, large open circles; Mg in H04, filled pentagons) are found in the left-bottom region with high [$\alpha$/Fe] and low [Mn/Fe].
ii) The thick disk stars (small open circles) populate the following region, [$\alpha$/Fe] $\sim 0.2-0.4$ and [Mn/Fe] $\sim -0.4$ to $-0.2$.
iii) The thin disks stars (small closed circles) occur at [$\alpha$/Fe] $\sim 0.1$ and [Mn/Fe] $\sim -0.1$, which are formed from the ISM largely enriched by SNe Ia.
In other words, it is possible to select thick disk stars only from the elemental abundance ratios.

The scatter should come from the variety in the SN II yields.
There is a larger scatter in H04 than C04, which may be partially caused by the observational error; the right-bottom star CS22952-015 is (-0.1, -0.34) in H06, but (0.36, -0.33) in C04 and A10. On the other side, the left-top star BS16085-050 is at (0.7, -0.01), which has to be checked. If it is real, the enrichment source may be faint SNe.

\section{Conclusions}

\begin{deluxetable}{lccccc}
\tablenum{3}
\tablecaption{\label{tab:summary}
Stellar Populations in Each Component.
}
\footnotesize
\tablewidth{0pt}
\tablehead{
Component & age & [(O, Mg, Si, S, Ca)/Fe] & [(Na, Al, Cu)/Fe] & [Zn/Fe] & [Mn/Fe]
}
\startdata
solar neighborhood & young & low & $\sim 0$ & $\sim 0$&  $\sim 0$ \\
bulge              & old & high & high & $\sim 0.2$ & low \\
thick disk         & old & high & $\sim 0$ & $\sim 0$ & low \\
\enddata
\end{deluxetable}

We present the chemodynamical simulations of a Milky Way-type galaxy using a self-consistent hydrodynamical code with supernova feedback and chemical enrichment. 
In our nucleosynthesis yields of core-collapse supernovae, the light curve and spectra fitting of individual supernova are used to estimate the mass of the progenitor, explosion energy, and ejected iron mass.
A large contribution from hypernovae is required from the observed abundance of Zn ([Zn/Fe] $\sim0$) especially at [Fe/H] $\ltsim -1$.
In our progenitor model of SNe Ia, based on the single degenerate scenario, the SN Ia lifetime distribution spans a range of $0.1-20$ Gyr with the double peaks at $\sim 0.1$ and $1$ Gyr.
Because of the metallicity effect of white dwarf winds, the SN Ia rate is very small at [Fe/H] $\ltsim -1$, which plays an important role in chemical evolution of galaxies.

In the simulated galaxy, the kinematical and chemical properties of the bulge, disk, and halo are consistent with the observations.
The bulge formed from the assembly of subgalaxies at $z \gtsim 3$; 80\% of bulge stars are older than $\sim 10$ Gyr, and 60\% have [O/Fe] $>0.3$.
The disk formed with constant star formation over $13$ Gyr; 50\% of solar-neighborhood stars are younger than $\sim 8$ Gyr, 80\% have [O/Fe] $<0.3$.
When we define the thick disk from kinematics, the thick disk stars tend to be older and have higher [$\alpha$/Fe] than the thin disk stars.
The formation timescale of the thick disk is $3-4$ Gyr.

Because the star formation history is different for different components, the age-metallicity relation and the metallicity distribution function are also different.
The age-metallicity relation shows a more rapid increase in the bulge than in the disk.
In both cases, the average metallicity does not show strong evolution at $t \gtsim 2$ Gyr, as in the observations.
The scatter is originated from the inhomogeneity of chemical enrichment in our chemodynamical model.
The observed metallicity distribution function is better reproduced with UV background radiation and hypernovae, but is still problematic.
The bulge wind induced by SNe Ia seems to be a good solution to reduce the numbers of metal-rich stars in the bulge and of metal-poor stars in the disk.

The difference in the chemical enrichment timescales results in the difference in the elemental abundance ratios, since different elements are produced by different supernovae with different timescales.
We also predict the frequency distribution of elemental abundance ratios as functions of time and location, which will be statistically compared with a large homogeneous sample from galactic archeology surveys such as HERMES, when they become available.

\begin{itemize}
\item
Because of the delayed enrichment of SNe Ia, $\alpha$ elements (O, Mg, Si, S, and Ca) show a plateau at [Fe/H] $\sim -1$, and then the decreasing trend against [Fe/H], where [Mn/Fe] also shows the increasing trend.
Odd-Z elements (Na, Al, and Cu) show the increasing trend at [Fe/H] $\ltsim -1$ because of the metallicity dependence of nucleosynthesis yields.
These are in excellent agreement with the available observations.

\item
In the bulge, the star formation timescale is so short that the [$\alpha$/Fe] plateau continues to [Fe/H] $\sim +0.3$.
Because of the smaller contribution from SNe Ia, the majority of stars shows high [$\alpha$/Fe] and low [Mn/Fe].
[(Na, Al, Cu, Zn)/Fe] are also high because of the high metallicity in the bulge.
\item
The stellar population of the thick disk is neither disk-like nor bulge-like as summarized in Table \ref{tab:summary}.
For thick disk stars, [$\alpha$/Fe] is higher, and [Mn/Fe] is lower than thin disk stars because of the short formation timescale.
However, [(Na, Al, Cu, Zn)/Fe] are lower than bulge stars because of the lower chemical enrichment efficiency.
This is because half of the thick disk stars have already formed in satellite galaxies before they accrete onto the disk, and the metals have been ejected from the satellite galaxies by the galactic winds.
\end{itemize}

\acknowledgments
We thank the National Astronomical Observatory of Japan for the GRAPE system, where most of the simulations in this paper are performed.
We also thank K. Nomoto, M. Mori, K. C. Freeman, G. Da Costa, and D. Yong for fruitful discussions.

\newpage

\begin{figure}
\center 
\includegraphics[width=11.5cm]{fig18.ps}
\caption{\label{fig:xfe-thick}
Same as Fig. \ref{fig:xfe} but for the thick disk.
The observational data (white dots) are taken from
\citet{pro00}, six-pointed stars;
\citet{ben04}, open triangles; 
and \citet{red06} and \citet{red08}, open circles.
}
\end{figure}

\begin{thebibliography}{}
\setlength{\parskip}{-2pt}

\bibitem[Anders \& Grevesse(1989)]{and89}
Anders, E., \& Grevesse, N. 1989, Geochimica et Cosmochimica Acta, 53, 197

\bibitem[Andrievsky et al.(2010)]{and10}
Andrievsky, S. M., Spite, M., Korotin, S. A., Spite, F., Bonifacio, P., Cayrel, R., Fran\c cois, P., \& Hill, V. 2010, \aap, 509, 88 

\bibitem[Andrievsky et al.(2007)]{and07}
Andrievsky, S. M., Spite, M., Korotin, S. A., Spite, F., Bonifacio, P., Cayrel, R., Hill, V., \& Fran\c cois, P. 2007, \aap, 464, 1081 

\bibitem[Andrievsky et al.(2008)]{and08}
Andrievsky, S. M., Spite, M., Korotin, S. A., Spite, F., Bonifacio, P., Cayrel, R., Hill, V., \& Fran\c cois, P. 2008, \aap, 481, 481 

\bibitem[Argast et al.(2002)]{arg02}
Argast, D., Samland, M., Thielemann, F.-K., \& Gerhard, O. E., 2002, \aap, 388, 842

\bibitem[Asplund et al.(2010)]{asp10}
Asplund, M., Grevesse, N., Sauval, A. J., \& Scott, P. 2009, \araa, 47, 481

\bibitem[Beers(2010)]{bee10}
Beers, T. 2010, in IAU Symposium 265 Chemical Abundances in the Universe: Connecting First Stars to Planets, eds. K. Cunha et al., p.453

\bibitem[Bensby et al.(2003)]{ben03}
Bensby, T., Feltzing, S., \& Lundstr\"{o}m, I. 2003, \aap, 410, 527

\bibitem[Bensby et al.(2004)]{ben04}
Bensby, T., Feltzing, S., \& Lundstr\"{o}m, I. 2004, \aap, 415, 155 

\bibitem[Bergemann \& Gehren(2008)]{ber08}
Bergemann, M., \& Gehren, T. 2008, \aap, 492, 823

\bibitem[Bergemann \& Gehren(2010b)]{ber10b}
Bergemann, M., \& Gehren, T. 2010, in IAU Symposium 265 Chemical Abundances in the Universe: Connecting First Stars to Planets, eds. K. Cunha et al., p.348

\bibitem[Bergemann et al.(2010)]{ber10}
Bergemann, M. Pickering, J. C., \& Gehren, T. 2010, \mnras, 401, 1334

\bibitem[Bertschinger(1995)]{ber95}
Bertschinger, E. 1995, http://arcturus.mit.edu/cosmics/

\bibitem[Bromm \& Larson(2004)]{bro04}
Bromm, V., \& Larson, R. B. 2004, \araa, 42, 79
	
\bibitem[Bruenn et al.(2009)]{bru09}
Bruenn, S. W., Mezzacappa, A., Hix, W. R., Blondin, J. M., Marronetti, P., Messer, O. E. B., Dirk, C. J., \& Yoshida, S. 2009, AIP Conference Proceedings, Vol. 1111, p. 593

\bibitem[Cayrel et al.(2004)]{cay04}
Cayrel, R., et al. 2004, \aap, 416, 1117

\bibitem[Chen et al.(2002)]{che02}
Chen, Y. Q., Nissen, P. E., Zhao, G., \& Asplund, M. 2002, \aap, 390, 225

\bibitem[Cunha et al.(2007)]{cun07}
Cunha, K., Sellgren, K., Smith, V. V.. Ramirez, S. V.. Blum, R. D., \& Terndrup, D. M., 2007, \apj, 669, 1011

\bibitem[Diemand, Kuhlen, \& Madau(2007)]{die07}
Diemand, J., Kuhlen, M., \& Madau, P. 2007, \apj, 657, 262

\bibitem[Edvardsson et al.(1993)]{edv93} 
Edvardsson, B., Andersen, J., Gustafsson, B., Lambert, D. L., Nissen, P. E.,
 \& Tomkin, J. 1993, \aap, 275, 101

\bibitem[Feltzing, Fohlman, \& Bensby(2007)]{fel07}
Feltzing, S., Fohlman, M., \& Bensby, T. 2007, \aap, 467, 665

\bibitem[Freeman et al.(2010)]{fre10}
Freeman, K. C., et al. 2010, in American Astronomical Society Meeting, Vol. 42, p.478

\bibitem[Fulbright(2000)]{ful00}
Fulbright, J. P. 2000, \aj, 120, 1841

\bibitem[Fulbright, McWilliam, \& Rich(2007)]{mcw04}
Fulbright, J. P., McWilliam, A., \& Rich, R. M. 2007, \apj, 661, 1152

\bibitem[Gratton et al.(2003)]{gra03}
Gratton, R. G., et al. 2003, \aap, 404, 187

\bibitem[Hachisu et al.(1999)]{hac99u} 
Hachisu, I., Kato, M., Nomoto, K., \& Umeda, H. 1999, \apj, 519, 314

\bibitem[Holmberg et al.(2007)]{hol07}
Holmberg, J., Nordstr\"{o}m, B., \& Andersen, J. 2007, \aap, 475, 519

\bibitem[Honda et al.(2004)]{hon04}
Honda, S., et al. 2004, \apj, 607, 474

\bibitem[Honma et al.(2007)]{hon07}
Honma, M., et al. 2007, \pasj, 59, 889

\bibitem[Hopkins et al.(2009)]{hop09}
Hopkins, P. F., et al. 2009, \mnras, 397, 802

\bibitem[Israelian et al.(1998)]{isr98}
Israelian, G., Garc{\'i}a-L{\'o}pez, R. J., \& Rebolo, R. 1998, \apj, 207, 805

\bibitem[Israelian \& Reboro(2001)]{isr01}
Israelian, G., \& Reboro, R. 2001, \apj, 557, L43

\bibitem[Iwamoto et al.(1999)]{iwa99}
Iwamoto, K., Brachwitz, F., Nomoto, K., Kishimoto, N., Umeda, H., Hix, W. R., \& Thielemann, F.-K. 1999, \apjs, 125, 439

\bibitem[Katz(1992)]{kat92}
Katz, N. 1992, \apj, 391, 502

\bibitem[Kauffmann \& White(1993)]{kau93} 
Kauffmann, G., \& White, S. D. M. 1993, \mnras, 261, 921

\bibitem[Kitaura, Janka \& Hillebrandt(2006)]{kit06}
Kitaura, F. S., Janka, H.-Th., \& Hillebrandt, W. 2006, \aap, 450, 345

\bibitem[Klypin, Zhao, \& Somerville(2002)]{kly02}
Klypin, A., Zhao, H., \& Somerville, R. S. 2002, \apj, 573, 597

\bibitem[Kobayashi(2004)]{kob04} 
Kobayashi, C., 2004, \mnras, 347, 740

\bibitem[Kobayashi(2005)]{kob05} 
Kobayashi, C., 2005, \mnras, 361, 1216

\bibitem[Kobayashi(2010)]{kob09b}
Kobayashi, C. 2010, in IAU Symposium 265 Chemical Abundances in the Universe: Connecting First Stars to Planets, eds. K. Cunha et al., p.336

\bibitem[Kobayashi et al.(2003)]{kob03} 
Kobayashi, C., Nakasato, N., \& Nomoto, K. 2003, in IAU Symposium 208 Astrophysical Supercomputing using Particle Simulations, 2001, eds. J. Makino \& P. Hut (Dordrecht: Kluwer), p.419

\bibitem[Kobayashi \& Nomoto(2009)]{kob09} 
Kobayashi, C., \& Nomoto, K. 2009, \apj, 707, 1466 (KN09)

\bibitem[Kobayashi, Karakas, \& Umeda(2010a)]{kob10a} 
Kobayashi, C., Karakas, I. A., \& Umeda, H. 2010a, \mnras, submitted

\bibitem[Kobayashi, Tominaga, \& Nomoto(2010b)]{kob10b} 
Kobayashi, C., Tominaga, N., \& Nomoto, K. 2010b, \apj, submitted, arXiv:1101.1227

\bibitem[Kobayashi et al.(2007)]{kob07} 
Kobayashi, C., Springel, V., \& White, S. D. M. 2007, \mnras, 376, 1465 (K07)

\bibitem[Kobayashi et al.(1998)]{kob98} 
Kobayashi, C., Tsujimoto, T., Nomoto, K., Hachisu, I, \& Kato, M. 1998, 
\apj, 503, L155

\bibitem[Kobayashi et al.(2006)]{kob06} 
Kobayashi, C., Umeda, H., Nomoto, K., Tominaga, N., \& Ohkubo, T. 2006, \apj, 653, 1145 (K06)

\bibitem[Koda et al.(2009)]{kod09}	
Koda, J., Milosavljevi{\'c}, M., \& Shapiro, P. R. 2009, \apj, 696, 254

\bibitem[Kodama \& Arimoto(1997)]{kod97} 
Kodama, T., \& Arimoto, N. 1997, \aap, 320, 41

\bibitem[Lecureur et al.(2007)]{lec07}
Lecureur, A., et al. 2007, \aap, 465, 799

\bibitem[Levine, Heiles, \& Blitz(2008)]{lev08}
Levine, E. S., Heiles, C., \& Blitz, L. 2008, \apj, 679, 1288

\bibitem[L{\'o}pez-Corredoira et al.(2002)]{lop02}
L{\'o}pez-Corredoira, M., Cabrera-Lavers, A., Garz{\'o}n, F., \& Hammersley, P. L. 2002, \aap, 394, 883

\bibitem[Macial, Lago, \& Costa(2006)]{mac06}
Macial, W. J., Lago, L. G., \& Costa, R. D. D. 2006, \aap, 453, 587

\bibitem[Maeda \& Nomoto(2003)]{mae03}
Maeda, K., \& Nomoto, K. 2003, \apj, 598, 1163

\bibitem[Mannucci et al.(2006)]{man06}
Mannucci, F., Della Valle, M., \& Panagia, N. 2006, \mnras, 370, 773

\bibitem[Marek \& Janka(2009)]{mar09}
Marek, A., \& Janka, H.-Th. 2009, \apj, 694, 664

\bibitem[McWilliam \& Rich(1994)]{mcw94}
McWilliam, A., \& Rich, R. M. 1994, \apjs, 91, 749

\bibitem[Merrifield(1992)]{mer92}
Merrifield, M. R. 1992, \aj, 103, 1552

\bibitem[Meynet \& Maeder(2002)]{mey02}
Meynet, G., \& Maeder, A. 2002, \aap, 390, 561

\bibitem[Moore et al.(1999)]{moo99}
Moore, B., Ghigna, S., Governato, F., Lake, G., Quinn, T., Stadel, J., \& Tozzi, P. 1999, \apj, 524, 19

\bibitem[Morrison(2010)]{mor10}
Morrison, H. 2010, talk at the conference on The Chemical Enrichment of the Milky Way Galaxy, at Ringberg Castle, Germany

\bibitem[Nakasato \& Nomoto(2003)]{nak03}
Nakasato, N., \& Nomoto, K. 2003, \apj, 588, 842

\bibitem[Navarro \& White(1993)]{nav93}
Navarro, J. F., \& White, S. D. M. 1993, \mnras, 265, 271

\bibitem[Nissen et al.(2007)]{nis07}
Nissen, P. E., Akerman, C., Asplund, M., Fabbian, D., Kerber, F., K\"aufl, H. U., \& Pettini, M. 2007, \aap, 469, 319

\bibitem[Nomoto(1982)]{nom82}
Nomoto, K. 1982, \apj, 253, 798

\bibitem[Nomoto \& Kondo(1991)]{nom91}
Nomoto, K., \& Kondo, Y. 1991, \apj, 367, L19

\bibitem[Nomoto et al.(2006)]{nom06}
Nomoto, K., Tominaga, N., Umeda, H., Kobayashi, C, \& Maeda, K. 2006, Nuclear Physics A, 777, 424

\bibitem[Nomoto et al.(1997a)]{nom97a}
Nomoto, K., et al. 1997a, Nuclear Physics, A616, 79c

\bibitem[Nomoto et al.(1997b)]{nom97}
Nomoto, K., et al. 1997b, Nuclear Physics, A621, 467c

\bibitem[Nordstr\"{o}m et al.(2004)]{nor04}
Nordstr\"{o}m, B., et al. 2004, \aap, 418, 989

\bibitem[Ohkubo et al.(2006)]{ohk06}
Ohkubo, T., Umeda, H., Maeda, K., Nomoto, K., Suzuki, T., Tsuruta, S., \& Rees, M. J. 2006, \apj, 645, 1352

\bibitem[Podsiadlowski et al.(2004)]{pod04}
Podsiadlowski, P., Mazzali, P. A., Nomoto, K., Lazzati, D., \& Cappellaro, E. 2004, \apj, 607, L17

\bibitem[Primas et al.(2000)]{pri00}
Primas, F., Reimers, D., Wisotzki, L., Reetz, J., Gehren, T., \& Beers, T. C. 2000, in The First Stars, ed. A. Weiss, T. Abel, \&, V. Hill (Berlin:Springer), 51

\bibitem[Prochaska et al.(2000)]{pro00}
Prochaska, J., Naumov, S. O., Carney, B, W,, McWilliam, A., \& Wolfe, A. M. 2000, \aj, 120, 2513

\bibitem[Reddy \& Lambert(2008)]{red08}
Reddy, B. E., \& Lambert, D. L. 2008, \mnras, 391, 95

\bibitem[Reddy, Lambert, \& Prieto(2006)]{red06}
Reddy, B. E., Lambert, D. L., \& Prieto, C. A. 2006, \mnras, 367, 1329

\bibitem[Reddy et al.(2003)]{red03}
Reddy, B. E., Tomkin, J., Lambert, D. L., \& Prieto, C. A. 2003, \mnras, 340, 304

\bibitem[Renzini et al.(1993)]{ren93}	
Renzini, A., Ciotti, L., D'Ercole, A., \& Pellegrini, S. 1993, \apj, 419, 52

\bibitem[Ruchti et al.(2010)]{ruc10}
Ruchti, G. R., et al. 2010, \apj, 721, L92

\bibitem[Sadler et al.(1996)]{sad96}
Sadler, E, M., Rich, R. M., \& Terndrup, D. M. 1996, \aj, 112, 171

\bibitem[Saito et al.(2009)]{sai09}
Saito, Y., Takada-Hidai, M., Honda, S., \& Takeda, Y. 2009, \pasj, 61, 549

\bibitem[S{\'a}nchez-Bl{\'a}zquez et al.(2009)]{san09}
S{\'a}nchez-Bl{\'a}zquez, P., Courty, S., Gibson, B. K., \& Brook, C. B. 2009, \mnras, 398, 591 

\bibitem[Scannapieco et al.(2009)]{sca09}
Scannapieco, C., White, S. D. M., Springel, V., \& Tissera, P. B. 2009, \mnras, 396, 696

\bibitem[Sommer-Larsen, G\"{o}lz, \& Portinari(2003)]{som03}
Sommer-Larsen, J., G\"{o}lz, M., \& Portinari, L. 2003, \apj, 596, 47

\bibitem[Steinmetz \& Navarro(1999)]{ste99}
Steinmetz, M., \& Navarro, J. F. 1999, \apj, 513, 555

\bibitem[Sullivan et al.(2006)]{sul06}
Sullivan, M., et al. 2006, \apj, 648, 868

\bibitem[Sutherland \& Dopita(1993)]{sut93}
Sutherland, R. S., \& Dopita, M. A. 1993, \apjs, 88, 235

\bibitem[Takada-Hidai et al.(2005)]{tak05}
Takada-Hidai, M., et al. 2005, \pasj, 57, 347

\bibitem[Thielemann et al.(1996)]{thi96}
Thielemann, F.-K., Nomoto, K., \& Hashimoto, M. 1996, \apj, 460, 408

\bibitem[Tinsley(1980)]{tin80}
Tinsley, B. M. 1980, Fundamentals of Cosmic Phisics Vol.5, p.287

\bibitem[Toth \& Ostriker(1992)]{tot92}
Toth, G., \& Ostriker, J. P. 1992, \apj, 389, 5

\bibitem[Tsujimoto et al.(1995)]{tsu95}
Tsujimoto, T., Nomoto, K., Yoshii, Y., Hashimoto, M., Yanagida, S., \&
 Thielemann, F.-K. 1995, \mnras, 277, 945

\bibitem[Warren et al.(1992)]{war92}
Warren, M. S., Quinn, P. J., Salmon, J. K., \& Zurek, W. H. 1992, \apj, 399, 405

\bibitem[Wyse \& Gilmore(1995)]{wys95}
Wyse, R. F. G., \& Gilmore, G. 1995, \aj, 110, 2771

\bibitem[Zoccali et al.(2008)]{zoc08}
Zoccali, M., et al. 2008, \aap, 486, 177

\end{thebibliography}
\end{document}